\documentclass[%
 aip,
jcp,
 amsmath,amssymb,floatfix,
 reprint,%
]{revtex4-1}
\newcommand{\dbtilde}[1]{\accentset{\approx}{#1}}
\usepackage{accents}
\usepackage{amsmath,color, tikz}
\usetikzlibrary{matrix}
\usepackage{amssymb}
\usepackage{amsfonts}
\usepackage{amsthm}
\usepackage{mathtools}
\usepackage{comment}
\usepackage{hyperref}
\usepackage{cleveref}
\usepackage{graphicx}
\usepackage{dcolumn}
\usepackage{bm}
\usepackage{xcolor}

\usepackage{physics}
\usepackage[utf8]{inputenc}
\usepackage[T1]{fontenc}
\usepackage{mathptmx}
\usepackage{etoolbox}
\usepackage{bbm}
\usepackage{chemformula}
\usepackage{amsmath}
\makeatletter
\def\@email#1#2{%
 \endgroup
 \patchcmd{\titleblock@produce}
  {\frontmatter@RRAPformat}
  {\frontmatter@RRAPformat{\produce@RRAP{*#1\href{mailto:#2}{#2}}}\frontmatter@RRAPformat}
  {}{}
}%
\makeatother
\begin{document}


\title[]{Two-dimensional IR--Raman spectroscopy of vibrational polaritons: Role of dipole surfaces}
\author{Xinwei Ji}
\affiliation{Department of Physics and Astronomy, University of Delaware, Newark, Delaware 19716, USA}

\author{Tomislav Begu\v{s}i\'c}
\affiliation{Institute of Physical and Theoretical Chemistry, University of W\"urzburg, 97074 W\"urzburg, Germany}
\email{tomislav.begusic@uni-wuerzburg.de}

\author{Tao E. Li}%
\affiliation{Department of Physics and Astronomy, University of Delaware, Newark, Delaware 19716, USA}
\email{taoeli@udel.edu}

\date{\today}

\begin{abstract}
Nonlinear spectroscopy provides a unique perspective to understand time-resolved molecular dynamics under vibrational strong coupling (VSC). Herein, equilibrium-nonequilibrium cavity molecular dynamics simulations are performed to compute the two-dimensional (2D) infrared-infrared-Raman (IIR) spectroscopy of liquid water under VSC. In conventional computational chemistry practices, accurate molecular spectra are often constructed by using an advanced molecular dipole or polarizability model to post-process  molecular dynamics trajectories evolved under a computationally efficient potential. By contrast, this work highlights the necessity of employing a consistent dipole surface model in both CavMD simulations and spectroscopic post-processing. While utilizing inconsistent dipole models only mildly influences the linear polariton spectrum, it severely distorts 2D spectra in wide frequency regions. With a consistent dipole-induced-dipole model, compared to the outside-cavity molecular 2D-IIR spectrum, the cavity 2D-IIR spectrum  splits the OH stretch band to a pair of polariton branches along only the IR (not Raman) axis, while fading molecular signals at other frequency regions. This work provides the foundation for employing direct CavMD simulations to construct 2D spectra of realistic molecules under VSC.
\end{abstract}

\maketitle

\section{Introduction}

Vibrational strong coupling (VSC) arises when a large number of molecular vibrations  interact  collectively with photonic normal modes in optical cavities. \cite{Shalabney2015,Long2015,Ebbesen2023,Wright2023}
Under this scenario, due to the formation of vibrational polaritons, experimental evidence suggests the modification of various molecular properties, including thermally-activated chemical reaction rates\cite{thomasGroundStateChemicalReactivity2016,Thomas2019_science,Ahn2023Science}, phase transitions\cite{sandeepClusterFormationPhase},  self-assembly\cite{hirai2021selective,sandeepManipulatingSelfAssemblyPhenyleneethynylenes2022,josephSupramolecularAssemblyConjugated2021},  polariton-excitation induced energy transfer\cite{xiangStateSelectivePolaritonDark2019,Xiang2020Science,Grafton2020,xiongMolecularVibrationalPolariton2023,Simpkins2023,Xiang2024}, and infrared (IR) photochemistry\cite{chenCavityenabledEnhancementUltrafast2022,yinOvercomingEnergyDisorder2025}. For investigating the underlying mechanism of VSC-modified chemistry, pump-probe and two-dimensional IR (2D-IR) spectroscopies have been employed to study ultrafast polariton dynamics at femtosecond time resolution. \cite{Dunkelberger2016,xiangStateSelectivePolaritonDark2019,Xiang2020Science,Grafton2020,xiongMolecularVibrationalPolariton2023,Simpkins2023,Xiang2024,chenCavityenabledEnhancementUltrafast2022,yinOvercomingEnergyDisorder2025,Pyles2024}

In parallel with the experimental exploration of VSC, a large volume of theoretical work has emerged during the past decade.\cite{Luk2017,flickAtomsMoleculesCavities2017,Hoffmann2018,trianaShapeElectricDipole2020,Botzung2020,Fregoni2022,Mandal2023ChemRev,YangCao2021,Yang2021QEDFT,Sokolovskii2022tmp,Riso2022,Suyabatmaz2023,Perez-Sanchez2023,Sukharev2023,Yu2024,Liu2024,Sharma2024,Wickramasinghe2025,Mondal2025} Among existing computational approaches for VSC, the classical cavity molecular dynamics (CavMD) scheme represents one economic yet powerful method for probing time-resolved polariton dynamics.\cite{liCavityMolecularDynamics2020,liCavityMolecularDynamics2021,liPolaritonRelaxationVibrational2022,jiSelectiveExcitationIRInactive2025,liQMMMModeling2023} In this approach, classical cavity modes are coupled to a large ensemble of condensed-phase molecules under electronic ground state. While classical propagation of coupled photonic-nuclear dynamics provides a computationally efficient method for simulating vibrational polaritons, path-integral extension of CavMD enables the semiclassical description of photonic and nuclear quantum effects on the same footing. \cite{liQuantumSimulationsVibrational2022}

Equipped with CavMD, many fundamental mechanisms of vibrational polaritons have been identified, including polariton-enhanced molecular nonlinear absorption\cite{liCavityMolecularDynamics2021,liPolaritonRelaxationVibrational2022}, polariton-induced selective excitation of solute molecules\cite{liEnergyefficientPathwaySelectively2022} or IR-inactive modes\cite{jiSelectiveExcitationIRInactive2025}, modification of IR photochemistry under polariton pumping\cite{liQMMMModeling2023}, and polariton-polariton scattering in multimode Fabry--P\'erot cavities\cite{liMesoscaleMolecularSimulations2024a}. Apart from our efforts on developing and utilizing CavMD, this approach is also gradually gaining popularity in both theoretical and experimental groups for studying VSC.\cite{lieberherrVibrationalStrongCoupling2023,bowlesLiquidWaterVibrational2025,yinOvercomingEnergyDisorder2025}

Despite these developments, significant experiment-theory disparity still remains: an end-to-end spectroscopic comparison between ultrafast experiments and CavMD simulations is not established. Currently, spectroscopic experiments interpret nonequilibrium polariton  dynamics from nonlinear or 2D-IR spectra, whereas CavMD directly identifies energy flow between different states from time-resolved simulations in the absence of 2D spectra \cite{liCavityMolecularDynamics2021,liPolaritonRelaxationVibrational2022,liEnergyefficientPathwaySelectively2022,liQMMMModeling2023,jiSelectiveExcitationIRInactive2025}. Hence, directly obtaining 2D spectra from CavMD simulations may futher facilitate our theoretical understanding and experimental interpretation of VSC dynamics.

\begin{figure}[htbp]
\centering
\includegraphics[width=1.0\linewidth]{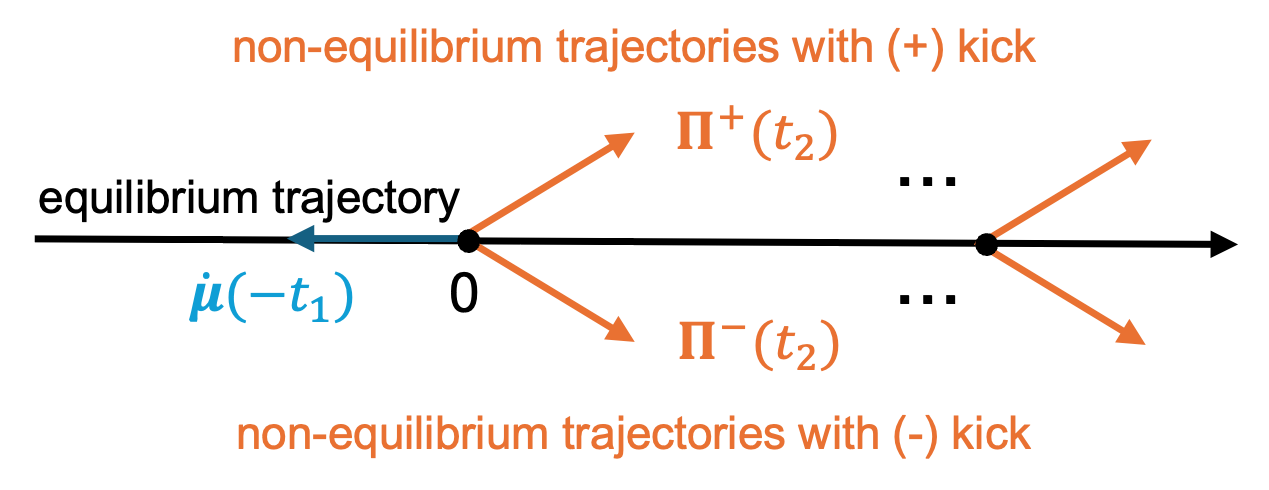}
\caption{\label{fig:2drf} \textbf{Scheme of the equilibrium-nonequilibrium MD approach for calculating the 2D-IIR response function defined in Eq. \eqref{eq:rfcm}.} Along the propagation of an equilibrium MD trajectory (black arrow), positive or negative delta pulse is applied at equilibrium molecular configurations (black nodes) to induce nonequilibrium molecular trajectories (yellow arrows). These nonequilibrium trajectories are used to evaluate the nonequilibrium polarizabilities  $\bm{\Pi}^{+}(t_2)$ and $\bm{\Pi}^{-}(t_2)$, whereas the time derivatives of the dipole vector $\dot{\boldsymbol{\mu}}$ are evaluated along the negative time  ($-t_1$) propagation of the equilibrium trajectory. }
\end{figure}

Outside the cavity, molecular dynamics (MD) simulations have been employed to model nonlinear or multidimensional spectroscopies. For example, 2D Raman and hybrid IR-Raman spectra can be simulated using equilibrium-nonequilibrium MD simulations.\cite{hasegawaCalculatingFifthorderRaman2006,itoCalculatingTwodimensionalTHzRamanTHz2014} In this approach, as illustrated in Fig. \ref{fig:2drf}, 2D spectra are obtained by estimating 2D response functions of molecular dipole moments at time $t_1$ correlated with polarizability tensors at time $t_2$. Such spectroscopies are powerful probes of mechanical and electrical anharmonic mode couplings.\cite{itoSimulatingTwoDimensional2016,Ciardi2019,Shalit2DRamanTHz2021,MousaviLowFrequencyAnharmonic2022,LinMapping2022,SeliyaOnSelectionRules2024} A recent computational work by one of us\cite{begusicTwodimensionalInfraredRamanSpectroscopy2023} identified features in the 2D IR-IR-Raman (2D-IIR) spectrum of liquid water that report on the the degree of tetrahedral arrangement between nearby water molecules. Same study also showed that the 2D-IIR spectra of liquid water remain largely unchanged when nuclear quantum effects are included via path-integral methods,\cite{begusicEquilibriumNonequilibriumRingpolymer2022}  suggesting the qualitative validity of performing classical simulations for capturing 2D-IIR spectra of liquid water.

Motivated by this work, here, we study 2D-IIR spectra of liquid water under VSC by performing equilibrium-nonequilibrium classical CavMD simulations. Outside the cavity, spectra are often constructed by post-processing computationally efficient MD trajectories using accurate and computationally demanding models of molecular dipole surfaces and polarizability tensors. \cite{Medders2015,Sommers2020,begusicTwodimensionalInfraredRamanSpectroscopy2023,Omodemi2025} Inside the cavity, we find that this outside-cavity recipe needs two major adjustments. (i) A brute-force post-processing of computationally efficient CavMD trajectories becomes insufficient for capturing 2D spectra under VSC: Because polaritons form due to the interaction between molecular dipole surfaces and photon modes, inconsistency of dipole surfaces used in CavMD simulations and post-processing may lead to strong numerical artifacts in describing the polariton signals in 2D spectra. In other words, for capturing accurate 2D spectra, advanced (and perhaps  computationally expensive) dipole surfaces must be directly employed when propagating CavMD. Apart from this issue, (ii) the form of 2D response functions inside the cavity also  differs from that outside the cavity. As the cavity IR spectra are dominated by the optical response of cavities rather than the confined molecules \cite{Carusotto2013}, cavity photon modes, rather than molecular dipoles, should enter 2D response functions for calculating VSC spectra. With these two adjustments, we systematically study the 2D-IIR spectra of liquid water under VSC and analyze the parameter dependence on the Rabi splitting and system size.

This paper is organized as follows. Sec. \ref{sec:theory} presents the theory of CavMD and equilibrium-nonequilibrium MD simulations for 2D-IIR spectroscopy. Sec. \ref{sec:simulation_details} outlines simulation details, and Sec. \ref{sec:implementation} provides the detailed implementation of the equilibrium-nonequilibrium CavMD  approach. We present the results in Sec. \ref{sec:result} and conclude in Sec. \ref{sec:conclusion}.

\section{Theory}\label{sec:theory}
In this section, we first briefly introduce the CavMD scheme. Next, we provide a detailed description of the two dipole surfaces employed in simulations. In the final two subsections, we outline the working equations for linear and nonlinear spectroscopies.

\subsection{Brief overview of CavMD}
The CavMD scheme builds upon the following dipole-gauge Hamiltonian under an electronic ground-state surface\cite{liCavityMolecularDynamics2020,liCavityMolecularDynamics2021,liPolaritonRelaxationVibrational2022}:
\begin{subequations}\label{eq:quantum_H_cavmd}
\begin{equation}\label{eq:hamiltonian}
\hat{H}^{\rm{G}}_{\rm{QED}} = \hat{H}^{\rm{G}}_{\rm{M}} + \hat{H}^{\rm{G}}_{\rm{F}} ,
\end{equation}
where $\hat{H}_{\mathrm{M}}^{\mathrm{G}}$ denotes the conventional ground-state molecular (kinetic + potential) Hamiltonian  in free space, and $\hat{H}_{\mathrm{F}}^{\mathrm{G}}$ represents the field-related Hamiltonian:
\begin{equation}\label{eq:hph}
\begin{aligned}
\hat{H}_{\mathrm{F}}^{\mathrm{G}}=&\sum_{k,\lambda}\frac{\hat{\widetilde{p}}_{k,\lambda}^{2}}{2 m_{k,\lambda}}+ \frac{1}{2}m_{k,\lambda}\omega_{k,\lambda}^{2}\left(\hat{\widetilde{q}}_{k,\lambda}+\frac{\varepsilon_{k,\lambda}}{m_{k,\lambda}\omega_{k,\lambda}^{2}}\sum_{n=1}^N\hat{d}_{\mathrm{ng},\lambda}\right)^{2} .
\end{aligned}
\end{equation}
\end{subequations}
Here, $\hat{\widetilde{p}} _{k, \lambda}, \hat{\widetilde{q}} _{k, \lambda}, \omega _{k, \lambda}, m_{k, \lambda}$ denote the momentum operator, position operator, frequency, and auxiliary mass for the cavity photon mode defined by the wave vector $\mathbf{k}$ ($k = |\mathbf{k}|$) and polarization unit vector $\bm{\xi}_{\lambda}$ (with $\mathbf{k} \cdot \bm{\xi}_{\lambda} = 0$). 
Each molecule interacts with the cavity photon modes via $\hat{d}_{\mathrm{ng},\lambda}$, the electronic ground-state dipole operator of the $n$-th molecule projected along the polarization direction $\bm{\xi}_{\lambda}$. The parameter $\varepsilon_{k,\lambda}$ characterizes the coupling strength between the cavity photon mode and each individual molecule, which is defined as 
\begin{equation}\label{eq:varepsilon_real_coupling}
    \varepsilon_{k,\lambda} \equiv \sqrt{\frac{m_{k,\lambda}\omega_{k,\lambda}^2}{\Omega\epsilon_0}} ,
\end{equation}
where $\Omega$ and $\epsilon_0$ represent the effective cavity volume and vacuum permittivity, respectively. In Eq. \eqref{eq:hph}, both the counter-rotating-wave terms and the dipole-self-energy term are included. Note that  the auxiliary mass $m_{k, \lambda}$ is introduced by rescaling the momentum and position operators of each cavity photon mode ($\hat{\widetilde{p}}_{k,\lambda}\equiv  \sqrt{m_{k\lambda}}\hat{p}_{k,\lambda}$ and $\hat{\widetilde{q}}_{k,\lambda}\equiv  \hat{q}_{k,\lambda} /\sqrt{m_{k\lambda}}$) and does not alter any VSC dynamics. 

Given the Hamiltonian in Eq. \eqref{eq:quantum_H_cavmd}, in the classical limit, CavMD propagates the following equations of motion:
\begin{subequations}\label{eq:eom}
\begin{align}
M_{nj}\ddot{\mathrm{R}}_{nj}&= \mathbf{F}_{nj}^{(0)}+\mathbf{F}_{nj}^{\mathrm{cav}},\label{eq:cavmd-eom-1}
\\
m_{k,\lambda}\ddot{\dbtilde{q}}_{k,\lambda} &= -m_{k,\lambda}\omega_{k,\lambda}^2\dbtilde{q}_{k,\lambda}-\widetilde{\varepsilon}_{k,\lambda} d_{\lambda} . \label{eq:cavmd-eom-2}
\end{align}
Here, $M_{nj}$ and $\mathbf{R}_{nj}$ denote the mass and position of the $j$-th nucleus in the $n$-th molecule, with $\mathbf{F}_{nj}^{(0)}$ representing the corresponding nuclear force in free space. The cavity force acting on each nucleus, $\mathbf{F}_{nj}^{\mathrm{cav}}$, will be introduced later in this section. 

Because it is computationally prohibitive to include the total number of molecules confined in the cavity ($N_{\rm{tot}}$) in simulations, the CavMD approach assumes that the whole molecular system can be represented by $N_{\rm{cell}}$ identical molecular simulation cells, each of which contains $N_{\mathrm{simu}}$ explicitly simulated molecules, with $N_{\rm{tot}} = N_{\rm{cell}} N_{\mathrm{simu}}$. \cite{liCavityMolecularDynamics2020} Under this approximation, each cavity photon mode interacts with the total dipole moment of explicitly simulated molecules, $d_{\lambda}\equiv \sum_{n=1}^{N_{\mathrm{simu}}}d_{ng,\lambda}$, and the corresponding effective light-matter coupling strength per molecule, $ \widetilde{\varepsilon}_{k,\lambda}$, becomes 
\begin{equation}\label{eq:varepsilon_effective_coupling}
    \widetilde{\varepsilon}_{k,\lambda}=\sqrt{N_{\rm{cell}}}\varepsilon_{k,\lambda} = \sqrt{\frac{N_{\rm{cell}} m_{k,\lambda}\omega_{k,\lambda}^2}{\Omega\epsilon_0}} ,
\end{equation}
with the real coupling strength $\varepsilon_{k,\lambda}$ defined in Eq. \eqref{eq:varepsilon_real_coupling}. Under this approximation, the coordinates of all cavity photon modes are also rescaled by $\dbtilde{q}_{k,\lambda}=\widetilde{q}_{k,\lambda}/\sqrt{N_{\rm{cell}}}$.

In Eq. \eqref{eq:cavmd-eom-1}, the cavity force acting on each nucleus is then given by:
\begin{equation}\label{eq:fcav}
\mathbf{F}_{nj}^{\mathrm{cav}}=-\sum_{k,\lambda}\biggl(\widetilde{\varepsilon}_{k,\lambda}\dbtilde{q}_{k,\lambda}+\frac{\widetilde{\varepsilon}_{k,\lambda}^2}{m_{k,\lambda}\omega_{k,\lambda}^2}d_{\lambda}\biggr)\frac{\partial d_{\lambda}}{\partial\mathbf{R}_{nj}}   .
\end{equation}
\end{subequations}
In this work, we adopt single-mode CavMD for practical simulations \cite{liCavityMolecularDynamics2020,liCavityMolecularDynamics2021,liPolaritonRelaxationVibrational2022}, i.e., only a single cavity mode is included in simulations. With the assumption that the cavity mirrors are placed along the $z$-direction, the cavity mode is then polarized  and coupled to the molecules along both the $x$- and $y$-directions (with $\lambda = x,y$). In this single-mode CavMD scheme, the long-wave approximation is taken, as the micrometer-scaled EM wavelength for IR vibrations exceeds the molecular length scale significantly. The multimode mesoscale CavMD approach \cite{liMesoscaleMolecularSimulations2024a} is not applied here. After all, most reported 2D VSC spectra do not involve cavity photon modes at different frequencies.

It is worth-noting that in CavMD, both nuclear forces and molecular dipole moments adopt the outside-cavity free-space forms. While these molecular properties may be significantly altered in small-volume nanocavities \cite{szidarovszkyEfficientFlexibleApproach2023,schnappingerMolecularPolarizabilityVibrational2025}, they remain largely unchanged in micrometer-scaled Fabry--P\'erot cavities \cite{Li2020Origin}. Thus, for using CavMD to explore cavity effects on molecular dynamics in the Fabry--P\'erot limit, it becomes reasonable to simply utilize the outside-cavity molecular properties in the calculations.

\subsection{Evaluating dipole surfaces}\label{sec:dipole}

The above CavMD equations of motion require three pieces of molecular information at each time step: the nuclear forces outside the cavity, $\mathbf{F}_{nj}^{(0)}$; the  total molecular dipole moment, $d_{\lambda}$; and the dipole derivatives with respect to nuclear coordinates, $\partial d_{\lambda}/\partial\mathbf{R}_{nj}$.  In this work, the nuclear forces outside the cavity, $\mathbf{F}_{nj}^{(0)}$, are always evaluated using the same force field --- the q-TIP4P/F water model.\cite{habershonCompetingQuantumEffects2009}  Two approaches, however, are employed for evaluating the molecular dipoles and their derivatives: (i) the fixed point-charge model, and (ii) the truncated dipole-induced-dipole (DID) model \cite{begusicTwodimensionalInfraredRamanSpectroscopy2023,hamm2DRamanTHzSpectroscopySensitive2014}. We note that more accurate potential and dipole surfaces are readily used in the literature, ranging from \textit{ab initio} methods\cite{HeydenAbInitioMolecularDynamics2010,WanRamanSpectraOfLiquidWater2013,MarsalekQuantumDynamicsAndSpectroscopy2017,RozsaAbInitioSpectroscopy2018,PestanaTheQuestForAccurate2018,CassoneAbInitioSpectroscopy2019,OhtoAccessingTheAccuracy2019} to fitted polarizable\cite{HasegawaAPolarizableWaterModel2011,hamm2DRamanTHzSpectroscopySensitive2014,meddersInfraredRamanSpectroscopy2015,itoEffectsIntermolecularCharge2016,sidlerEfficientWaterForce2018} or machine-learning force fields.\cite{GasteggerMachineLearningMolecularDynamics2017,MorawietzTheInterplayOfStructure2018,KapilInexpensiveModelingOfQuantumDynamics2020,SchienbeinSpectroscopyFromMachineLearning2023,InoueIsUnifiedUnderstanding2023,KapilFirstPrinciplesSpectroscopy2024} In this work, the q-TIP4P/F model provides a computationally efficient flexible water force field, whereas the selected dipole models include one that is strictly linear in nuclear coordinates (based on point charges) and another that accounts for nonlinearities in the dipole and polarizability surfaces (a property essential for spectroscopic observables).

\subsubsection{The fixed point-charge model}
In the fixed point-charge model, the total dipole moment $d_{\lambda}$ is calculated by $d_{\lambda}=\sum_{n}^{N_{\text{simu}}}d_{ng,\lambda}$, where the dipole moment for each molecule, $d_{ng,\lambda}$, is defined as
\begin{equation}\label{eq:dipole_fixed_charge}
d_{ng,\lambda}=\sum_jQ_{nj}R_{nj,\lambda} .
\end{equation}
Here, $Q_{nj}$ denotes the pre-assigned fixed partial charge of the $j$-th atom in the $n$-th molecule, and $R_{nj,\lambda} \equiv \mathbf{R}_{nj}\cdot \boldsymbol{\xi}_{\lambda}$. With Eq. \eqref{eq:dipole_fixed_charge}, the dipole derivatives can be obtained as:
\begin{equation}\label{eq:dipole_derivatives_fixed_charge}
\frac{\partial d_{\lambda}}{\partial R_{nj,\lambda'}}=Q_{nj} \delta_{\lambda'\lambda},
\end{equation}
where $\delta_{\lambda,\lambda'}$ denotes the Kronecker delta function. When the direction $\lambda = \lambda'$ (with $\lambda, \lambda' \in \{x,y,z\}$), Eq. \eqref{eq:dipole_derivatives_fixed_charge} equals to $Q_{nj}$; otherwise it becomes zero.

\subsubsection{The truncated dipole-induced-dipole model}
In the DID model, the total dipole moment vector $\bm{\mu}$ and polarizability tensor $\bm{\Pi}$ are given by\cite{begusicTwodimensionalInfraredRamanSpectroscopy2023,hamm2DRamanTHzSpectroscopySensitive2014}:
\begin{subequations}
\begin{align}
\label{eq:mu_did}
\bm{\mu}&=\sum_{n}\bm{\mu}_{n}+\bm{\mu}^{\mathrm{ind}},\\
\bm{\Pi}&=\sum_{n}\bm{\Pi}_{n}+\bm{\Pi}^{\mathrm{ind}},
\end{align}
\end{subequations}
where $\bm{\mu}_n$ and $\bm{\Pi}_n$ represent the permanent dipole vector and polarizability tensor of the $n$-th molecule, respectively.
The permanent dipole vector $\bm{\mu}_n$ is evaluated by assuming fixed point charges for the atoms, in a manner analogous to Eq. \eqref{eq:dipole_fixed_charge}.
Similarly, calculating the permanent polarizability of the $n$-th molecule, $\bm{\Pi}_n$, requires  the coordinates of the same molecule only \cite{avilaInitioDipolePolarizability2005}. 
Apart from these permanent contributions, the induced dipole moment $\bm{\mu}^{\mathrm{ind}}$ and polarizability $\bm{\Pi}^{\mathrm{ind}}$ are modeled as 
\begin{subequations}\label{eq:ind}
\begin{align}
\bm{\mu}^{\mathrm{ind}}&=\sum_{n \neq m} \bm{\Pi}_n\cdot\mathbf{E}_{nm},\label{eq:muind}\\
\bm{\Pi}^{\mathrm{ind}}&=-\sum_{n \neq m} \bm{\Pi}_n\cdot\mathbf{T}_{nm}\cdot\bm{\Pi}_m,\label{eq:piind}
\end{align}
\end{subequations}
where the subscripts $n,m$ denote the $n$-th and $m$-th molecules, respectively. $\mathbf{E}_{nm}$ represents the local electric field produced by molecule $m$ acting on the oxygen atom of molecule $n$, and $\bm{T}_{nm}$ represents 
the $3 \times 3$ dipole-dipole interaction tensor\cite{itoNotesSimulatingTwodimensional2015,itoEffectsIntermolecularCharge2016}.

With the above analytical definition of the total dipole vector $\boldsymbol{\mu}$, the corresponding dipole derivatives are evaluated using the chain rule.

\subsection{Linear spectroscopies}
The linear IR spectrum outside the cavity is calculated by evaluating the molecular dipole autocorrelation function\cite{liCavityMolecularDynamics2020,habershonComparisonPathIntegral2008,habershonCompetingQuantumEffects2009,MorawietzTheInterplayOfStructure2018}:
\begin{equation}
    I^{\mathrm{IR}}_{\rm m}(\omega) \propto \omega^2\int_{-\infty}^{\infty} \mathrm{d}t e^{-i \omega t} \langle\bm{\mu}(0)\bm{\mu}(t)\rangle,\label{eq:1d_ir}
\end{equation}
where $\bm{\mu}(t)$ denotes the total dipole vector of the molecular system at time $t$.

With single-mode CavMD, because the molecular system is coupled to the cavity photon mode in two possible polarization directions ($x$- and $y$-directions), the molecular IR spectrum is calculated by
\begin{equation}
    I^{\mathrm{IR}}_{\rm{m}}(\omega) \propto  \sum_{i=x,y} \omega^2 \int_{-\infty}^{\infty} \mathrm{d}t e^{-i \omega t} \langle(\bm{\mu} (0) \cdot \mathbf{e}_i) (\bm{\mu}(t) \cdot \mathbf{e}_i)\rangle .\label{eq:1d_ir_cavity}
\end{equation}
Here, $\mathbf{e}_i$ denotes the unit vector along the $i$-th dimension.
As the $z$-component of the molecular dipole vector is decoupled from the cavity, this component is excluded from spectroscopic calculations.

In experiments, the measured IR spectrum under VSC is dominated by the optical response of the cavity rather than the molecules. \cite{Carusotto2013} In other words, denoting the effective dipole moment of the optical transitions in the cavity as $\mu_{k\lambda} = Q_{k\lambda} \dbtilde{q}_{k\lambda}$, where $Q_{k\lambda}$ is the associated effective charge, the magnitude of $\mu_{k\lambda}$  significantly surpasses that of molecules ($\bm{\mu}$), or $|\mu_{k\lambda}| \gg |\bm{\mu}| $.  Consequently, for the strongly coupled light-matter system, the experimentally observed IR spectrum under VSC should stem from the effective dipole moment of the cavity mode rather than the molecules. Therefore, in a manner
similar to Eq. \eqref{eq:1d_ir_cavity}, we compute the linear IR spectrum of the cavity photon modes as\cite{liMesoscaleMolecularSimulations2024a}
\begin{equation}\label{eq:IR_cavity_linear}
    I^{\mathrm{IR}}_{\rm c}(\omega) \propto  \sum_{\lambda=x,y} \omega^2 \int_{-\infty}^{\infty} \mathrm{d}t e^{-i \omega t} \langle \dbtilde{q}_{\lambda}(0) \dbtilde{q}_{\lambda}(t)\rangle .
\end{equation}
Note that in the current CavMD simulations, 
because only a single cavity mode (polarized along both the $x$- and $y$-directions) is taken into account, the $k$-dependence in photonic coordinates is omitted for simplicity.

After introducing the linear IR spectrum, we also compute the linear Raman spectrum of the molecular system. While in experiments Raman spectroscopy is measured with a highly off-resonant visible light source, in molecular simulations it is frequently calculated using the response functions of the molecular polarizability tensor $\bm{\Pi}$ evaluated in the absence of external light. \cite{MorawietzTheInterplayOfStructure2018} Specifically, following Ref. \citenum{MorawietzTheInterplayOfStructure2018}, Raman spectra of molecules can be calculated by Fourier transforming the polarizability tensor:
\begin{subequations}\label{eq:1d-raman}
\begin{align}
I^{\rm{iso\text{-}Raman}}_{\rm m}(\omega)&\propto \frac{\omega}{1-e^{-\beta\hbar\omega}}\int_{-\infty}^{\infty} \mathrm{d}t e^{-i \omega t} \langle\bm{\alpha}(0)\bm{\alpha}(t)\rangle,\label{eq:iso_raman}\\
I^{\rm{aniso\text{-}Raman}}_{\rm m}(\omega)&\propto \frac{\omega}{1-e^{-\beta\hbar\omega}}\int_{-\infty}^{\infty} \mathrm{d}t e^{-i \omega t} \langle\mathrm{Tr}[\bm{\beta}(0)\bm{\beta}(t)]\rangle . \label{eq:aniso_raman}
\end{align}
\end{subequations}
Here, $\beta = 1/k_{\rm B} T$, where $k_{\rm B}$ and $T$ denote the Boltzmann constant and temperature, respectively.  The isotropic Raman spectrum [Eq. \eqref{eq:iso_raman}] is calculated by evaluating $\bm{\alpha}(t)=\mathrm{Tr}[\bm{\Pi}(t)]/3$, whereas the anisotropic Raman spectrum [Eq. \eqref{eq:aniso_raman}] is computed using $\bm{\beta}(t)=\bm{\Pi}(t)-\mathbbm{1}(\mathrm{Tr}[\bm{\Pi}(t)]/3)$, with $\mathbbm{1}$ being the identity matrix. 

Eq. \eqref{eq:1d-raman} should be valid for evaluating the molecular Raman spectra both inside and outside the cavity. This is because,
inside the cavity, the highly off-resonant Raman excitation does not directly measure the cavity signals in the IR domain. Instead, it is the molecular response which dominates the Raman signal.

\subsection{2D-IIR spectroscopy}

Outside the cavity, the 2D-IIR response function is defined as \cite{itoCalculatingTwodimensionalTHzRamanTHz2014}
\begin{equation}\label{eq:rfqm}
R_{\rm m}(t_1,t_2)=-\frac{1}{\hbar^2}\mathrm{Tr}\{[[\hat{\bm{\Pi}}(t_1+t_2),\hat{\bm{\mu}}(t_1)],\hat{\bm{\mu}}(0)]\hat{\rho}\} .
\end{equation}
Here, $\hat{\bm{\Pi}}, \hat{\bm{\mu}}, \hat{\rho}$ represent the polarizability operator, dipole operator, and thermal density operator of the molecular system, respectively. Via the equilibrium-nonequilibrium MD approach\cite{hasegawaCalculatingFifthorderRaman2006,itoCalculatingTwodimensionalTHzRamanTHz2014,itoNotesSimulatingTwodimensional2015,sunHybridEquilibriumnonequilibriumMolecular2019}, the IIR response function can be  approximately evaluated by
\begin{equation}\label{eq:rfcm}
R^{\mathrm{MD}}_{\rm m}(t_1,t_2)=\frac{\beta}{\varepsilon}\langle[\bm{\Pi}^+(t_2)-\bm{\Pi}^-(t_2)]\dot{\bm{\mu}}(-t_1)\rangle .
\end{equation}
Here,  the trace in Eq. \eqref{eq:rfqm} is replaced by the classical thermal average over MD trajectories ($\langle\cdots\rangle$); $\bm{\Pi}$ and $\dot{\bm{\mu}}$ denote the classical molecular polarizability tensor and time derivative of the dipole vector, respectively.

A practical numerical scheme for computing the classical  IIR response function [Eq. \eqref{eq:rfcm}] is  depicted in Fig. \ref{fig:2drf}. Starting from molecular configuration sampled from the thermal equilibrium distribution (a node along the black line), $\bm{\Pi}^{+}(t_2)$ or $\bm{\Pi}^{-}(t_2)$ represents the nonequilibrium  polarizability tensor at later times ($t_2$) after sending a positive ($+$) or negative ($-$) delta pulse to perturb the initial molecular momenta. In detail, after the delta pulse polarized along the $x$-direction, the molecular momenta are displaced from the equilibrium momenta $\bm{p}$ to 
\begin{equation}\label{eq:p_change}
    \bm{p}^{\pm}=\bm{p}\pm\frac{\varepsilon}{2}\nabla\bm{\mu}\cdot\mathbf{e}_x ,
\end{equation}
where $\varepsilon$ denotes the magnitude of the perturbation by the electric field, and the quantity $\nabla\bm{\mu}$ represents the dipole derivatives with respect to molecular coordinates. The quantity $\dot{\bm{\mu}}(-t_1)$ is evaluated along the negative time propagation ($-t_1$) of the equilibrium molecular trajectory (the opposite direction of the black arrow in Fig. \ref{fig:2drf}).
Note that in Eq. \eqref{eq:rfcm}, while the response function is in general a multidimensional tensor, because liquid water is an isotropic system, we take only the $xx$-component of the polarizability tensor and the $x$-direction of the dipole vector for  practical calculations. Inside the cavity, the $x$-component is also directly coupled to the cavity, so polaritonic signals can be recovered.

After obtaining the response function in the time domain, the 2D-IIR spectra are computed through a double sine transform\cite{hammTwodimensionalRamanterahertzSpectroscopyWater2012,itoEffectsIntermolecularCharge2016}
\begin{equation}\label{eq:dsin}
R_{\rm m}(\omega_1,\omega_2)=\int_0^\infty\int_0^\infty R_{\rm m}^{\rm{MD}}(t_1,t_2)\sin(\omega_1t_1)\sin(\omega_2t_2) \mathrm{d} t_1 \mathrm{d} t_2.
\end{equation}
where $R_{\rm m}^{\rm{MD}}(t_1,t_2)$ has been defined in Eq. \eqref{eq:rfcm}.

Inside the cavity, because the cavity typically responds to the IR probe much more strongly than the confined molecules, the following modified 2D-IIR response function may more faithfully report the experimental measurement:
\begin{equation}\label{eq:rfcm_c}
R^{\mathrm{MD}}_{\rm c}(t_1,t_2)=\frac{\beta}{\varepsilon}\langle[\bm{\Pi}^+(t_2)-\bm{\Pi}^-(t_2)]\dot{\dbtilde{q}}_{\lambda=x}({-t_1})\rangle .
\end{equation}
Compared to Eq. \eqref{eq:rfcm}, here the photonic coordinate is used to replace the molecular dipole. The corresponding 2D spectra are evaluated analogously to Eq. \eqref{eq:dsin}. It should be noted that the same delta pulse, polarized along the $x$-direction, as well as the displacement of dipole moments described in Eq. \eqref{eq:rfcm}, are applied to compute the cavity 2D-IIR spectra.

\section{Simulation Details}
\label{sec:simulation_details}

All spectroscopic data were obtained by simulating liquid water with classical MD. The non-polarizable q-TIP4P/F force field \cite{habershonCompetingQuantumEffects2009} was employed to calculate nuclear forces of the \ch{H2O} molecules. For intermolecular interactions, the cutoff of Lennard--Jones potential was set to $r_{\rm{cut}}=4.137$ \AA, and Ewald summation was applied to evaluate the long-range Coulomb interactions. A cubic simulation cell with length $a=12.42$ \AA \ containing $N_{\rm{simu}} = 64$ \ch{H2O} molecules was used, corresponding to the density $\rho=0.998$ g/cm$^3$ at temperature $T=280$ K. The initial water geometry was generated by PACKMOL\cite{martinezPACKMOLPackageBuilding2009}.  For simulations involving enlarged water systems, both the density and temperature of liquid water were maintained the same.

\begin{figure}[htbp]
\centering
\includegraphics[width=1.0\linewidth]{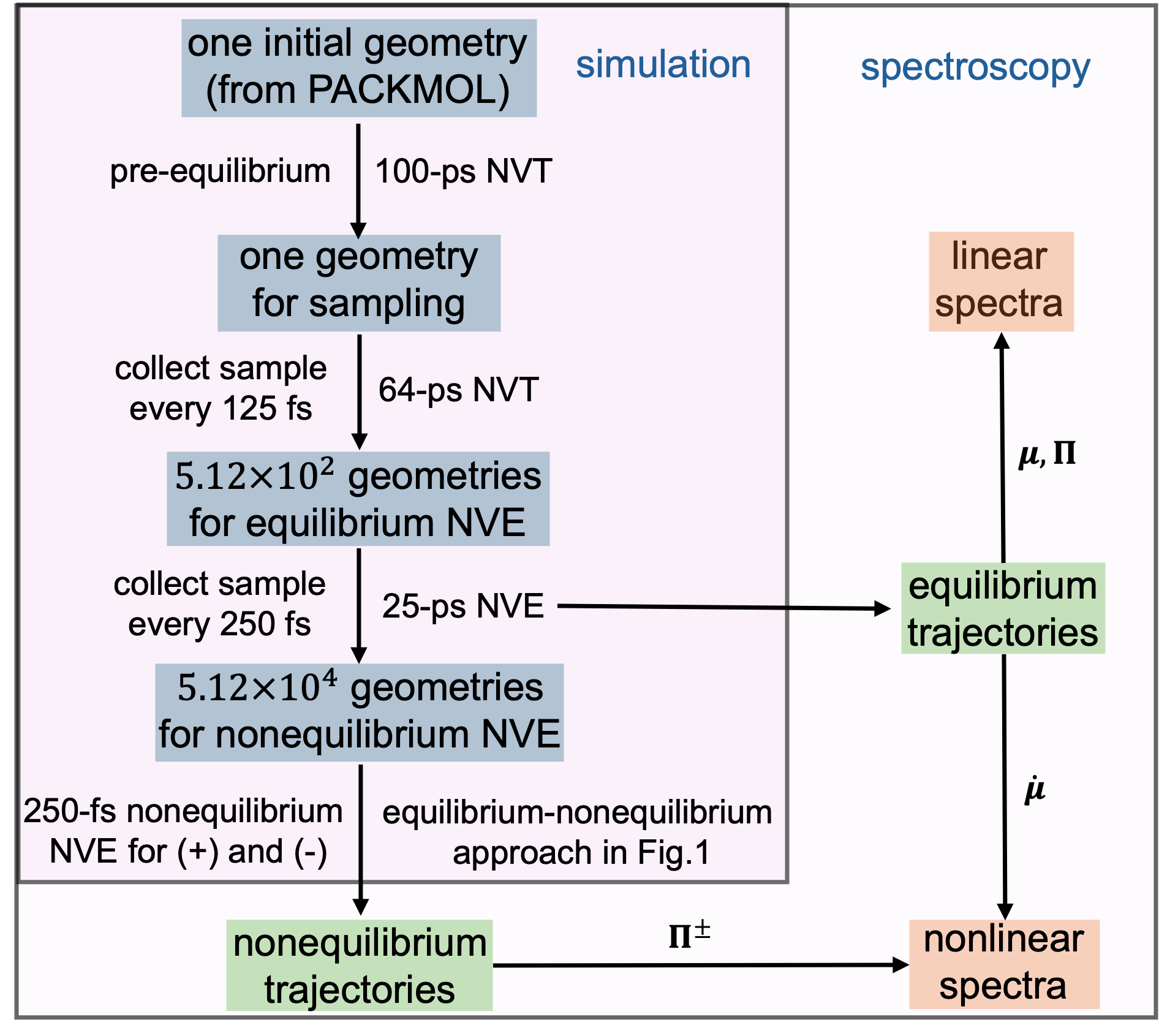}
\caption{\label{fig:workflow} \textbf{General workflow for evaluating linear and nonlinear spectra with direct equilibrium-nonequilibrium MD or CavMD simulations.} See Sec. \ref{sec:simulation_details} for the detailed description.}
\end{figure}

Outside the cavity, we largely followed  Ref. \citenum{begusicTwodimensionalInfraredRamanSpectroscopy2023} for computing linear and nonlinear spectroscopies.  As outlined in Fig. \ref{fig:workflow},  the initial water geometry was pre-equilibrated by performing a 100-ps simulation under the canonical (constant number, volume, and temperature, or NVT) ensemble at 280 K. The Langevin thermostat, with a Langevin friction lifetime of $\tau = 100$ fs, was applied during the NVT simulation. The time step of MD simulations was set to $\Delta t = 0.25$ fs. The final water geometry after the  pre-equilibration was used for an additional 64-ps NVT simulation, with the initial velocities resampled following a Maxwell--Boltzmann distribution at 280 K. During this subsequent NVT simulation, the water geometry was saved every 125 fs, producing 512  equilibrated water geometries. Starting from each of these 512 geometries,  a separate 10-ps NVT simulation is conducted to ensure the system reaching thermal equilibrium, followed by a subsequent simulation under the microcanonical (constant number, volume, and energy, or NVE) ensemble  for 25 ps.

During the 25-ps NVE simulations, the dipoles and polarizabilities were evaluated every 1 fs for plotting the linear spectroscopies defined in Eqs. \eqref{eq:1d_ir} and \eqref{eq:1d-raman}. Every 250 fs during each 25-ps NVE simulation, the positions and momenta of the water configuration were saved for preparing the nonequilibrium NVE simulations, thus producing $5.12\times 10^{4}$ initial configurations for nonequilibrium simulations. At the beginning of each nonequilibrium simulation, the momenta of all nuclei experienced the perturbation from a delta pulse polarized along the $x$-axis; see also the diagram depicted in Fig. \ref{fig:2drf}. It was worth noting that the dipole derivatives corresponding to the DID dipole model were always used for the delta-pulse perturbation [c.f. Eq. \eqref{eq:p_change}]. \footnote{If the dipole derivatives corresponding to the fixed point-charge dipole model [Eq. \eqref{eq:dipole_derivatives_fixed_charge}] were used to couple to the delta electric pulse, the resulting 2D-IIR spectra were not meaningfully modified.}  The magnitude of the delta electric pulse was set to $\varepsilon = 0.1$ a.u., which corresponded to an electric field amplitude of $E_0 = \varepsilon/\Delta t \approx 0.5$ V/\AA. The pulse amplitude was set to an either positive ($+$) or negative ($-$) sign, representing the positive or negative perturbation.  Each nonequilibrium simulation was performed for 250 fs. After both the equilibrium and nonequilibrium NVE simulations, the classical 2D-IIR response function were calculated via Eq. \eqref{eq:rfcm}, using the $xx$-component of the nonequilibrium polarizabilities, $\Pi^{\pm}_{xx}$, and the time derivative of the $x$-component of equilibrium dipole, $\dot{{\mu}}_x$. To examine convergence of the 2D spectra, when the molecular number was set to $N_{\rm{simu}}=64$, the whole procedure depicted in Fig. \ref{fig:workflow} was repeated for ten times. Since both the linear and nonlinear spectra had already converged without repeating the simulations for ten times, for simulations at other molecular sizes ($N_{\rm{simu}}=32$ and 128), this additional repeating was omitted.

Inside the cavity, water molecules were coupled to the cavity photon mode at frequency $\omega_{\rm{c}}=3550$ cm$^{-1}$ along both the $x$- and $y$-polarization directions. Two effective coupling strengths $\widetilde{\varepsilon} =4\times10^{-4}$ a.u. and $7\times10^{-4}$ a.u. were employed in CavMD simulations. All the other simulation details largely resembled those outside the cavity. It is worth noting, however, that during NVT simulations, the Langevin thermostat was applied to both the atoms and cavity photons  with the friction lifetime $\tau = 100$ fs. For exploring the size dependence of the spectra signals, the molecular number was changed from $N_{\rm{simu}}=64$ to $N_{\rm{simu}}=32$ and 128, respectively.  During the change of the molecular system size, the observed Rabi splitting was fixed the same by maintaining 
$\widetilde{\varepsilon}\cdot\sqrt{N_{\rm{simu}}/64} = 4\times10^{-4}$ a.u. or $7\times10^{-4}$ a.u., respectively.\cite{liCavityMolecularDynamics2020} 
For polariton spectra calculated in this manuscript, the cavity loss was frequently assumed to be zero; we validated this approximation using the calculation in Fig. \ref{fig:2d-ir-raman}d, which enforced a cavity lifetime of 10 ps.

\begin{figure*}
\centering
\includegraphics[width=0.9\linewidth]{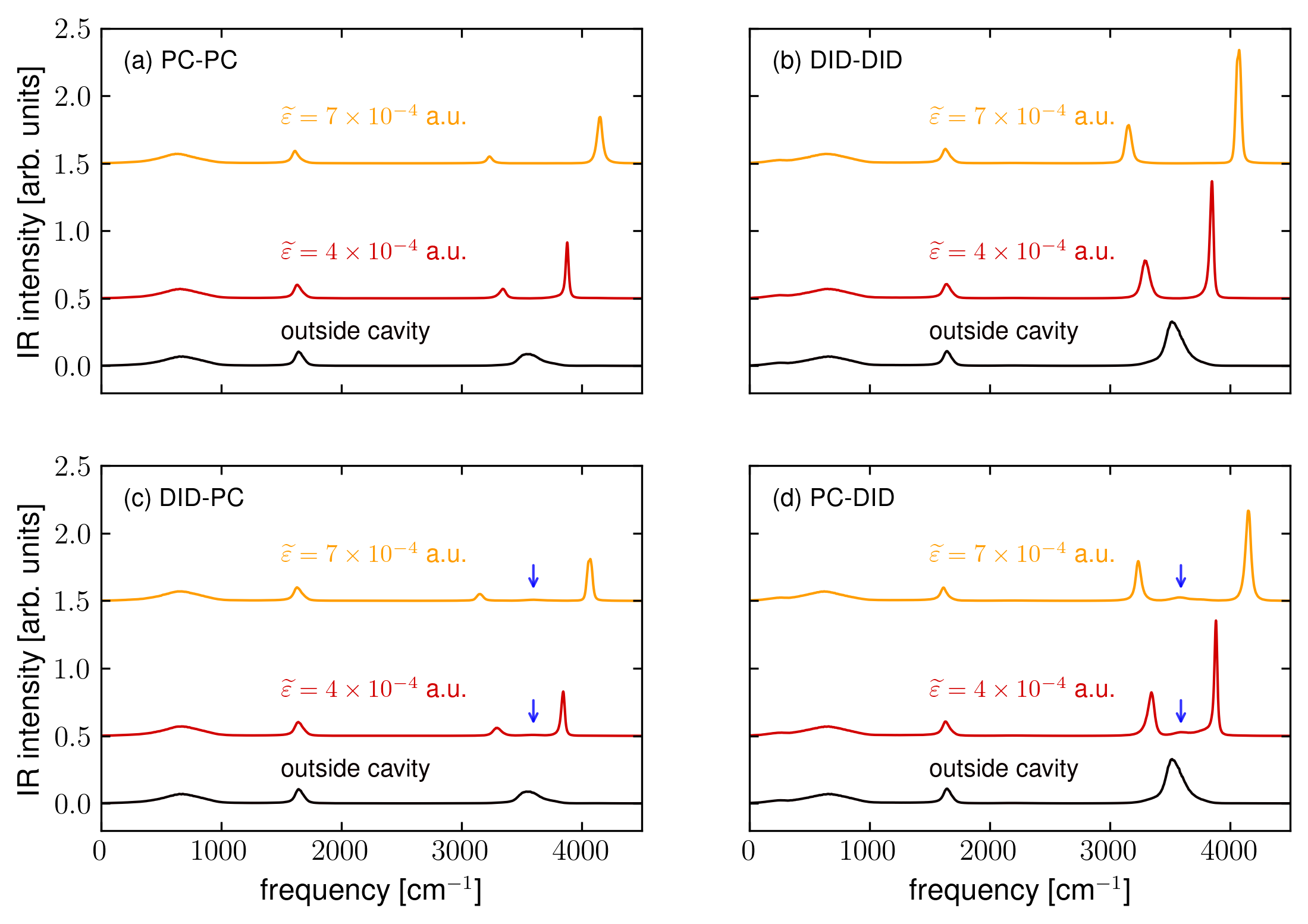}
\caption{\label{fig:1d-ir} \textbf{Simulated linear IR spectra of liquid water by evaluating molecular dipole autocorrelation functions.} In each panel, the outside-cavity spectrum (black) is compared against that under VSC with the effective light-matter coupling strength $\widetilde{\varepsilon}=4 \times 10^{-4}$ a.u. (red) or $7 \times 10^{-4}$ a.u. (orange). For CavMD propagation and spectroscopic calculations, both the fixed point-charge (PC) or the DID dipole model are employed: (a) PC in both CavMD propagation and spectroscopic calculations; (b) DID in both CavMD propagation and spectroscopic calculations; (c) DID in CavMD propagation and PC in spectroscopic calculations; (d) PC in CavMD propagation and DID in spectroscopic calculations. Each blue arrow indicates a weak middle peak between the LP and UP, arising from the inconsistency in the dipole models employed in CavMD simulations versus  spectroscopic calculations. For parameters, the cavity frequency is set to $\omega_{\rm{c}} = 3550$ cm$^{-1}$, and  the explicitly simulated water system contains $N_{\rm{simu}}=64$ molecules. }
\end{figure*}

\section{Implementation details}\label{sec:implementation}

Although this study simulated only  liquid water, we provided a general-purpose implementation for propagating equilibrium-nonequilibrium CavMD in a modified version of i-PI 3.0\cite{litmanIPI30Flexible2024}. 

In detail, coupling molecules to the cavity was supported by the  \texttt{FFCavPhSocket} interface in the i-PI 3.0 code.
This interface concurrently connected to two molecular drivers on the fly via a socket interface: (i) the LAMMPS MD driver for evaluating nuclear forces outside the cavity, and (ii) the DID water dipole driver for evaluating the values of molecular dipole vectors and polarizability tensors.

When the fixed point-charge dipole model was used to propagate CavMD, the \texttt{FFCavPhSocket} interface employed the fixed partial charges pre-assigned in the input to construct molecular dipoles and their geometric gradients [c.f. Eqs. \eqref{eq:dipole_fixed_charge} and \eqref{eq:dipole_derivatives_fixed_charge}] on the fly. Then, at each time step, this interface combined the dipole information, nuclear forces outside the cavity, and photonic coordinates, to calculate the nuclear and photonic forces in the cavity.  

When the DID dipole model was utilized to propagate CavMD, the \texttt{FFCavPhSocket} interface 
directly acquired the dipole vector and its geometric derivatives from the DID dipole driver at every time step to construct nuclear and photonic forces in the cavity, thus enabling the CavMD propagation
with accurate dipole surfaces. This flexible computation architecture allows CavMD simulations to use a wide range of dipole surfaces without further modifying the i-PI source code.

\begin{figure}[htbp]
\centering
\includegraphics[width=1.0\linewidth]{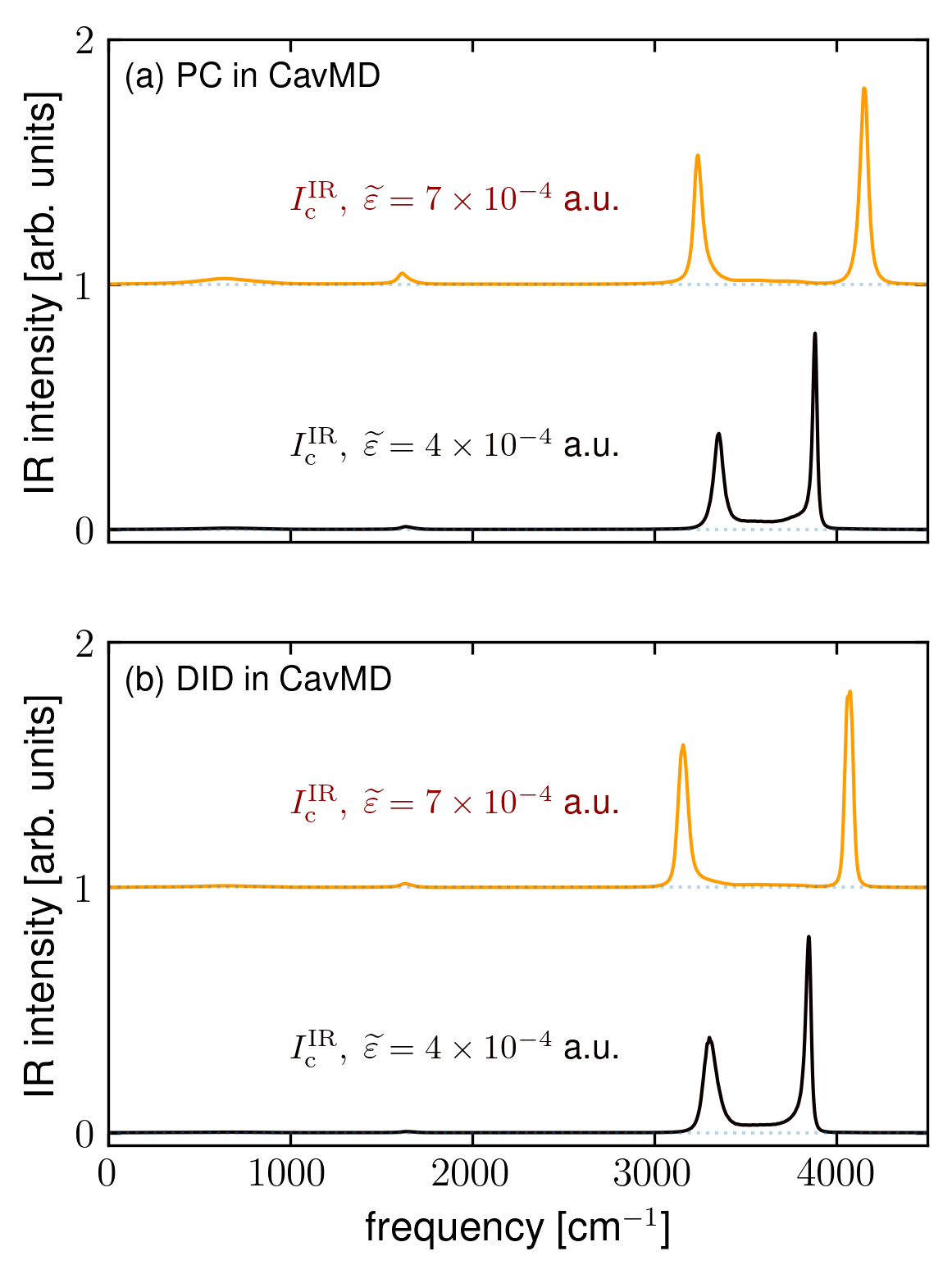}
\caption{\label{fig:1d-qc}\textbf{Simulated linear cavity IR spectra for liquid water under VSC.} The CavMD propagation employs the (a) fixed point-charge model or (b) DID dipole model. In each panel, the linear cavity IR spectra is plotted according to Eq. \eqref{eq:IR_cavity_linear} under two effective coupling strengths: $\widetilde{\varepsilon}=4 \times 10^{-4}$ a.u. (red) and $7 \times 10^{-4}$ a.u. (orange).  Beneath each cavity lineshape, a horizontal baseline (dotted gray) guides the visualization. All simulation parameters remain the same as those in Fig. \ref{fig:1d-ir}.}
\end{figure}

\section{Result}\label{sec:result}
After introducing the simulation and implementation details, we now investigate both the linear and 2D-IIR spectra of liquid water inside versus outside the cavity.

\subsection{Linear IR spectra}

We start with linear IR spectra of liquid water.
Fig. \ref{fig:1d-ir} plots the linear IR spectra by evaluating the Fourier transform of the molecular dipole autocorrelation function [Eqs. \eqref{eq:1d_ir} and \eqref{eq:1d_ir_cavity}]. Because both the CavMD propagation and linear spectral calculations rely on the molecular dipole information, depending on whether the dipole surface is described by the fixed point-charge model or the DID model, four different method combinations  become available for preparing the molecular linear IR spectrum. 

For instance, Fig. \ref{fig:1d-ir}a demonstrates the molecular linear IR spectra when both the CavMD propagation and spectroscopic calculations use the fixed point-charge model (denoted as PC-PC), with the outside-cavity spectrum (black) compared against the inside-cavity results under two different effective light-matter coupling strengths:
$\widetilde{\varepsilon}= 4\times10^{-4}$ a.u. (red) and $7\times10^{-4}$ a.u. (orange). For both the coupling strengths, because the cavity photon frequency ($\omega_{\text{c}}=3550$ cm$^{-1}$) is set at resonance with the \ch{O-H} stretch band, the wide \ch{O-H} stretch band near 3550 cm$^{-1}$ is split into a pair of lower polariton and upper polariton (LP and UP) peaks inside the cavity. In agreement with previous CavMD work, both the LP and UP possess linewidths much smaller than that of the bare \ch{O-H} stretch band outside the cavity. This linewidth difference suggests that the polariton states do not inherit the inhomogeneous linewidth broadening in the molecular transition \cite{Houdre1996,Long2015}.

Fig. \ref{fig:1d-ir}b provides the analogous polariton spectra when both the CavMD propagation and the spectroscopic calculation employ the DID dipole model (denoted as DID-DID).  Compared to the outside-cavity spectrum (black line) in Fig. \ref{fig:1d-ir}a, here the intensity of water \ch{O-H} stretch band is significantly enhanced, demonstrating improved agreement with the \textit{ab initio} reference in Ref. \citenum{meddersInfraredRamanSpectroscopy2015}. Following the same trend, the polariton intensities under VSC (red and orange lines) are also amplified compared to those in Fig. \ref{fig:1d-ir}a. 

More interestingly, when the CavMD propagation and the spectroscopic calculation adopt different models for the dipole surface (DID-PC or PC-DID), as shown in Figs. \ref{fig:1d-ir}c and Figs. \ref{fig:1d-ir}d, a weak middle peak (blue arrow) always emerges between the LP and UP. As the polariton spectra in Fig. \ref{fig:1d-ir}b, which employs the more accurate DID dipole model in both CavMD propagation and spectroscopic calculations, do not contain this middle peak, we must attribute this weak middle peak here to a computational anomaly instead of real experimental features. The physical origin of this anomaly can be understood as follows: During the CavMD propagation, the polariton states stem from the hybridization between the cavity photon mode and the molecular bright state, the latter being defined as the symmetric superposition of local molecular dipoles evaluated under one level of theory (or basis).  Then, if a different dipole model is used in spectroscopic post-processing, the new total dipole moment now represents a different combination of molecular coordinates and may not equal exactly to the bright state involved in the CavMD propagation --- it may instead contain a small contribution from the dark modes (the asymmetric superposition of local molecular dipoles) defined in the old basis. In short, it is this  dark-state contribution that yields the small middle peak between the two polariton states. A more quantitative analysis is further provided in the Appendix.

\begin{figure*}[htbp]
\centering
\includegraphics[width=0.9\linewidth]{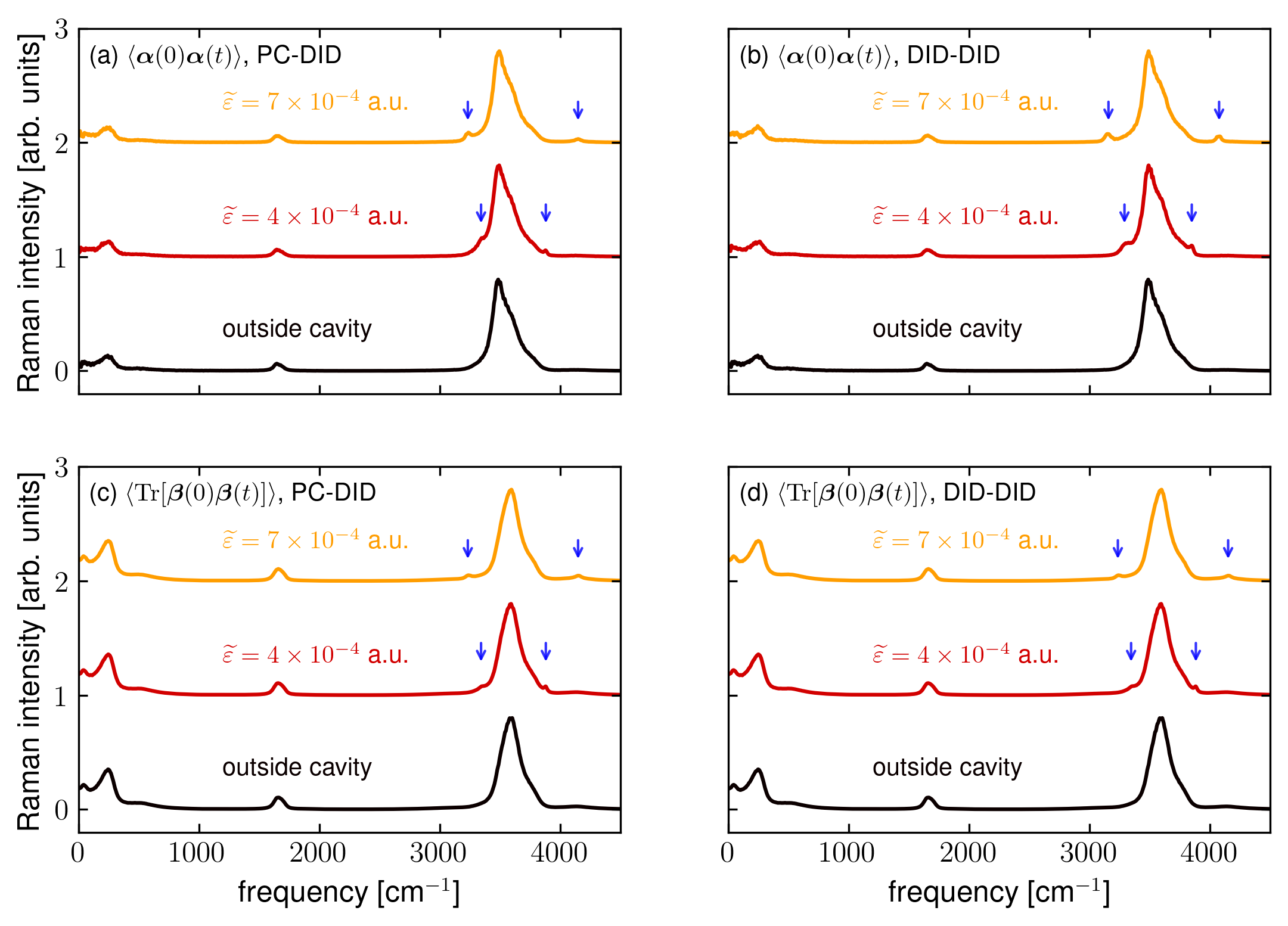}
\caption{\label{fig:1d-raman}
\textbf{Simulated linear  Raman spectra for liquid water inside versus outside the cavity.}  Both the isotropic (top panel) and anisotropic (bottom panel) Raman spectra are plotted. The CavMD propagation employs the (a,c) fixed point-charge (PC) model or (b,d) DID dipole model, while the spectroscopic post-processing always uses the DID model. In each part, the outside-cavity spectrum (black) is compared against that under VSC when the effective coupling strength is set to $\widetilde{\varepsilon}=4 \times 10^{-4}$ a.u. (red) or $7 \times 10^{-4}$ a.u. (orange). All simulation parameters remain the same as those in Fig. \ref{fig:1d-ir}. For the linshapes inside the cavity, blue arrows indicate the weak contributions from the LP and UP states, as only $N_{\rm{simu}}=64$ \ch{H2O} molecules are explicitly simulated. }
\end{figure*}

Beyond the molecular dipole spectra in Fig. \ref{fig:1d-ir},  an  experimentally  more relevant approach for examining VSC is to probe the cavity IR spectra. After all, for molecules confined in optical cavities, the cavity responds to the external IR measurement much more strongly than the confined molecules. Thus, we further plot the cavity IR spectra by Fourier transforming the photonic coordinate autocorrelation function  inside the cavity [Eq. \eqref{eq:IR_cavity_linear}]. 

As shown in Fig. \ref{fig:1d-qc}, the cavity IR spectra exhibit strong signals only in the polariton regions under the two effective light-matter coupling strengths: $\widetilde{\varepsilon}=4\times10^{-4}$ a.u. (brown) and $7\times10^{-4}$ a.u. (yellow). Here, the CavMD propagation employs either  the fixed point-charge dipole model (part a) or the DID dipole model (part b). In both cases, compared to the strongly asymmetric UP and LP peaks in the molecular linear IR spectra [Fig. \ref{fig:1d-ir}], the relative intensities of the  UP and LP peaks become more balanced. It is also worth noting that no middle peak exists between the polariton states in the cavity IR spectra. After all, the cavity coordinate used in spectroscopic calculations always synchronizes with the molecular dipole moment in CavMD propagation, thus involving no post-processing with a different dipole model.

\subsection{Raman spectra}

Having investigated linear IR spectra, we now study Raman spectra of liquid water. 
Fig. \ref{fig:1d-raman} shows the isotropic [Eq.~\eqref{eq:iso_raman}, upper panels] and anisotropic [Eq.~\eqref{eq:aniso_raman}, bottom panels] Raman spectra of liquid water inside versus outside the cavity, computed by Fourier transforming the polarizability tensor autocorrelation function. As the Raman signals always require the DID dipole model in spectroscopic post-processing, we further compare the spectra when the CavMD propagation employs the fixed point-charge  (left panels) or DID dipole model (right panels). In both cases, the Raman spectra inside the cavity (red and orange lines) largely resemble those outside the cavity (black line), apart from the appearance of small side polariton peaks (denoted by blue arrows). In agreement with previous studies \cite{DelPino2015,Takele2020,Ahn2021}, because Raman spectroscopy involves highly off-resonant excitation of the strong coupling system, it directly measures the behavior of individual molecules instead of the molecular bright state or cavity modes. As such, polariton signals become negligible here even when only $N_{\rm{simu}}=64$ molecules are explicitly included in the simulation. Very interestingly, the polariton signals in the anisotropic Raman spectra are weaker than those in the isotropic spectra.

\begin{figure*}[htbp]
\centering
\includegraphics[width=0.9\linewidth]{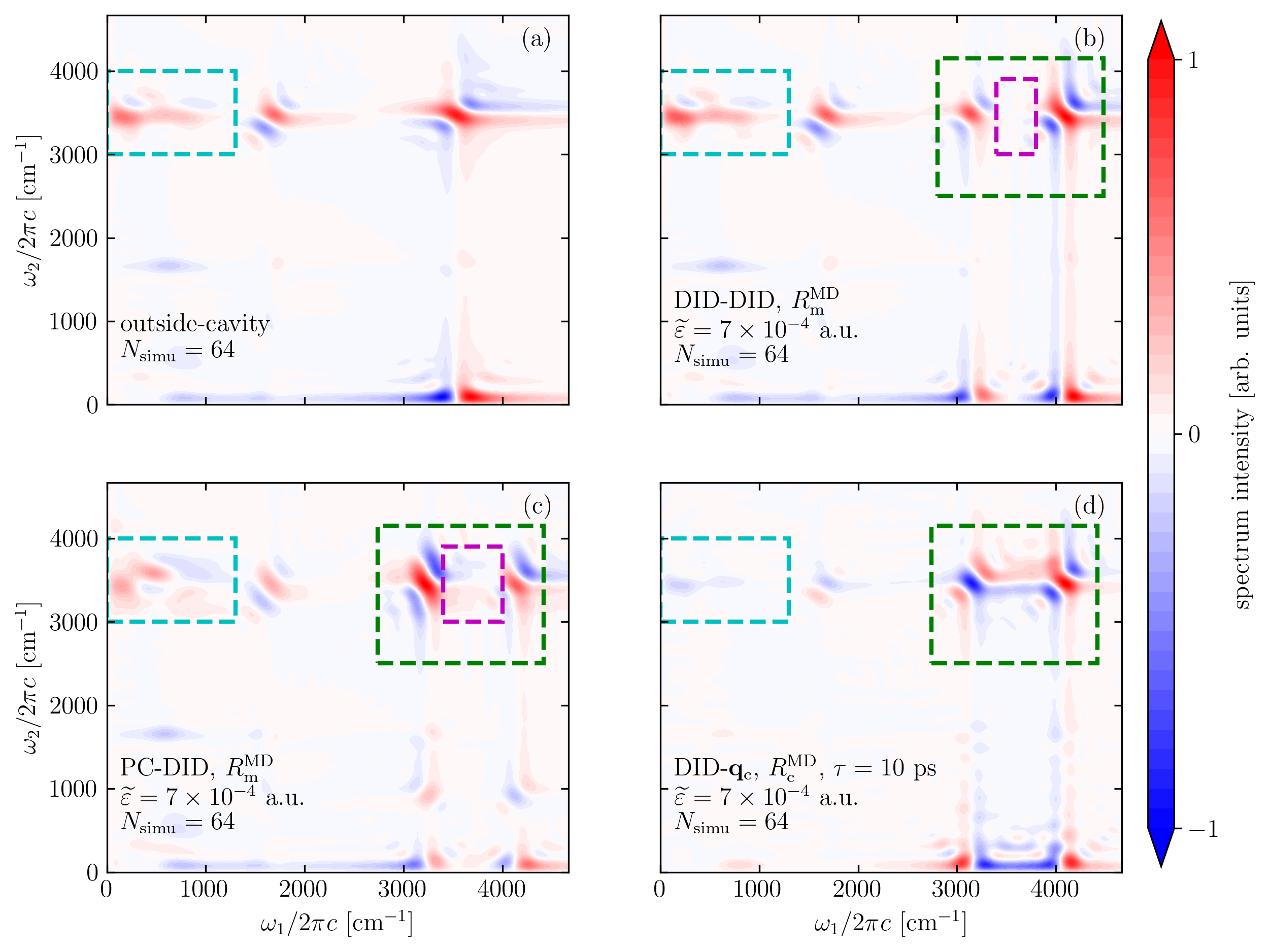}
\caption{\label{fig:2d-ir-raman}\textbf{Simulated 2D-IIR spectra of liquid water inside versus outside the cavity.} Four different scenarios are compared: (a) outside the cavity; molecular 2D-IIR spectra under VSC when the CavMD propagation utilizes (b) the DID dipole model or (c) the fixed point-charge (PC) model; and (d) cavity 2D-IIR spectrum when the CavMD propagation applies the DID model. In all cases, spectroscopic post-processing uses the DID dipole model. Employing inconsistent dipole surfaces in CavMD propagation and spectroscopic calculations (part c)  not only distorts the THz-IR-visible (TIRV, cyan rectangle) region outside the cavity, but also creates significant middle peaks (purple rectangle) within the the polariton signals (green rectangle). Compared to the molecular 2D-IIR spectra in part b, signals outside the polariton frequency window becomes strongly reduced in the cavity 2D-IIR spectrum (part d).}
\end{figure*}

\subsection{2D-IIR spectra}

Understanding linear IR and Raman spectra under VSC provides a foundation for the comprehension of 2D-IIR spectra. Following Ref. \citenum{begusicTwodimensionalInfraredRamanSpectroscopy2023}, Fig. \ref{fig:2d-ir-raman}a reproduces the 2D-IIR spectra of liquid water outside the cavity, computed via a double-sine transform of the IIR response function [Eqs. \eqref{eq:rfcm} and \eqref{eq:dsin}]. Here, the $x$- and $y$-axes, labeled with $\omega_1$ and $\omega_2$, correspond to the molecular dipole (IR) and  polarizability (Raman) signals, respectively. The former is evaluated using equilibrium MD trajectories, while the latter is computed through nonequilibrium MD dynamics. In consistency with Ref. \citenum{begusicTwodimensionalInfraredRamanSpectroscopy2023}, both the dipole and polarizability signals are evaluated using the DID dipole model by post-processing MD trajectories. In particular, 
the detailed structure of the THz-IR-Raman region (also called the THz-IR-visible\cite{grechkoCouplingIntraIntermolecular2018,vietzeDistinguishingDifferentExcitation2021,SeliyaExtractingTheSampleResponse2023}, or TIRV, region, labeled by the cyan rectangle) appears identical to that in Ref. \citenum{begusicTwodimensionalInfraredRamanSpectroscopy2023}. In the TIRV region, the strength of the negative signal (blue) around  $\omega_1 = 250$ cm$^{-1}$ and $\omega_2 =$ 3600 cm$^{-1}$ correlates with the tetrahedrality of the microscopic arrangement in \ch{H2O} molecules \cite{begusicTwodimensionalInfraredRamanSpectroscopy2023}.

Analogous to the 2D-IIR spectra outside the cavity, Fig. \ref{fig:2d-ir-raman}b plots the 2D-IIR spectra under VSC by evaluating the double-sine transform of the molecular  IIR response function [$R_{\rm{m}}^{\rm{MD}}$, Eqs. \eqref{eq:rfcm} and \eqref{eq:dsin}]. Here, the effective light-matter coupling strength is set to $\widetilde{\varepsilon}=7\times10^{-4}$ a.u., and the DID dipole model is employed in both CavMD simulations and spectroscopic post-processing. Importantly, under VSC the high-frequency IR-Raman region (green rectangle)  is split to a pair of LP and UP states only along the $x$-axis (IR probe). In agreement with linear Raman spectra (Fig. \ref{fig:1d-raman}), no meaningful polariton peak splitting can be observed along the $y$-axis (Raman probe). Additionally, due to the relative large Rabi splitting here, no measurable signal exists between the two polariton stripes along the $x$-axis (highlighted with purple rectangle). The TIRV region (cyan rectangle) also largely replicates the outside-cavity structure in part a. This consistency is within our expectation: The TIRV region reveals the anharmonic coupling between the low-frequency and high-frequency vibrational modes; although the low-frequency vibrations ($\omega_1$) are probed using IR light, these modes stay far off-resonant with respect to the IR cavity frequency; additionally, the high-frequency vibrations ($\omega_2$) are probed using Raman spectroscopy, which remains insensitive to the polariton formation (as in Fig. \ref{fig:1d-raman}).

We then switch the dipole model used in the CavMD propagation from the DID model to the fixed point-charge model, while keeping the employment of the DID model in spectroscopic calculations. Very interestingly, as shown in Fig. \ref{fig:2d-ir-raman}c, the molecular 2D-IIR spectra now becomes significantly distorted. In the high-frequency IR-Raman region (green rectangle), the UP stripe now becomes less intense than the LP one, an opposite trend compared to that in Fig. \ref{fig:2d-ir-raman}b. Moreover, between the LP and UP stripes along the $x$-axis, strong signals emerge at the middle (purple rectangle). The TIRV region (cyan rectangle) is also significantly modified --- the negative signal around  $\omega_1 = 250$ cm$^{-1}$ and $\omega_2 =$ 3600 cm$^{-1}$ in Figs. \ref{fig:2d-ir-raman}a,b, which indicates the tetrahedrality of water structures, now completely disappears. These three qualitative changes in the 2D spectra should correlate to the  inconsistency in the employed dipole models between CavMD propagation (using the fixed point-charge model) and spectroscopic post-processing (using the DID model). Obviously, this inconsistency brings more pronounced errors in 2D spectra than the 1D spectra.

While molecular 2D-IIR spectrum inside the cavity [Eq. \eqref{eq:rfcm}] provides direct comparison to that outside the cavity, because the cavity responds to the IR probe more significantly than the confined molecules, we further compute the cavity 2D-IIR spectrum in Fig. \ref{fig:2d-ir-raman}d using Eq. \eqref{eq:rfcm_c}. Here, both the CavMD propagation and spectroscopic post-processing utilize the DID dipole model. For better representing the experimental limit during productive calculations, instead of applying the NVE ensemble (Fig. \ref{fig:workflow}), we apply a weak Langevin thermostat on both molecules and photons with lifetime $\tau=10$ ps, representing a cavity loss lifetime at this value. Compared to the corresponding molecular 2D-IIR spectra (Fig. \ref{fig:2d-ir-raman}b), here the signs of the peaks flip, suggesting the opposite phase correlation between the molecular dipole and cavity mode. Additionally, comparing Fig. \ref{fig:2d-ir-raman}d against Fig. \ref{fig:2d-ir-raman}b,  the polariton signals (green rectangle) dominate the 2D spectra, whereas other spectroscopic regions, such as the TIRV region  (cyan rectangle), become strongly suppressed. 

\begin{figure*}[htbp]
\centering
\includegraphics[width=0.9\linewidth]{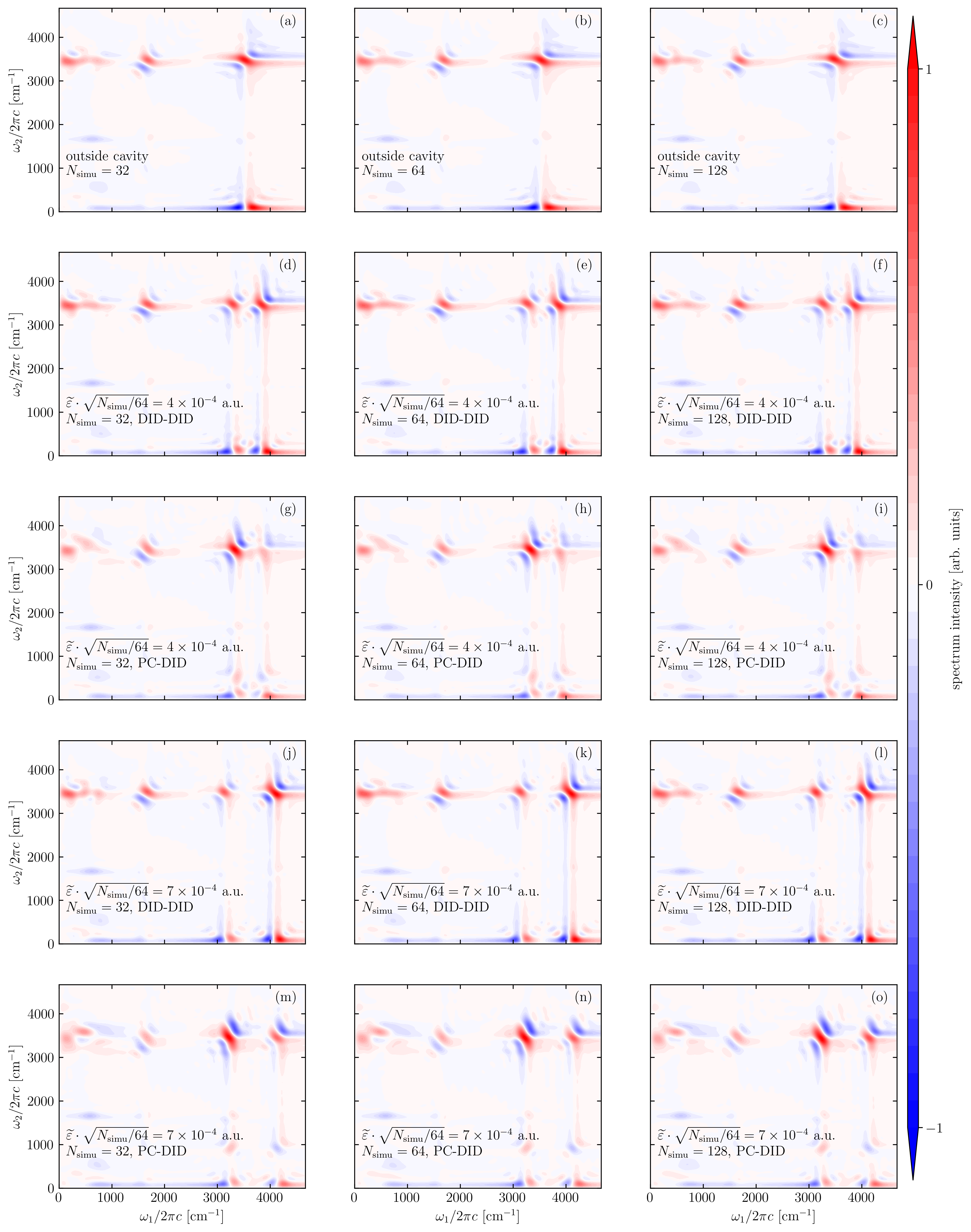}
\caption{\label{fig:2d_size_denpendence}\textbf{Simulated molecular 2D-IIR spectra of liquid water under varied molecular sizes.} (a-c) Spectra outside the cavity when the simulated molecular number is set to $N_{\rm{simu}}=32, 64$, and 128, respectively. Corresponding VSC spectra when the CavMD propagation uses the (d-f)  DID dipole model  or (g-i) fixed point-charge dipole model when the effective light-matter coupling strength is set to  $\widetilde{\varepsilon}  \sqrt{N_{\rm{simu}}/64} = 4\times 10^{-4}$ a.u. (j-o) Analogous VSC spectra at a larger effective coupling strength with $\widetilde{\varepsilon}  \sqrt{N_{\rm{simu}}/64} = 7\times 10^{-4}$ a.u.}
\end{figure*}

\subsection{System size dependence}


To end our discussion, we further report the molecular system size dependence of the 2D-IIR spectra under the constraint of fixed Rabi splitting. This study becomes crucial for understanding whether the above results with $N_{\rm{simu}}=64$ explicitly simulated \ch{H2O} molecules may correspond to the macroscopic experimental limit. In practice, fixing the Rabi splitting is achieved by varying the effective light-matter coupling strength $\widetilde{\varepsilon} \propto 1/ \sqrt{N_{\rm{simu}}}$. \cite{liCavityMolecularDynamics2020}

Similar to the molecular 2D-IIR spectra outside the cavity in Fig. \ref{fig:2d-ir-raman}a, the top row of Fig. \ref{fig:2d_size_denpendence} (parts a-c) demonstrates the molecular 2D-IIR spectra outside the cavity using $N_{\rm{simu}}=32, 64,$ and 128 explicitly simulated molecules, respectively. The maximum magnitude of the spectra intensity has been normalized to unity across the figures. Obviously, the 2D spectra remain largely invariant under different molecular system size. 

Then, inside the cavity, the second row of Fig. \ref{fig:2d_size_denpendence} (parts d-f) plots the molecular 2D-IIR spectra at varied system sizes given $\widetilde{\varepsilon}  \sqrt{N_{\rm{simu}}/64} = 4\times 10^{-4}$ a.u. Here, both the CavMD propagation and spectroscopic post-processing utilize the DID dipole model. Clearly, these spectra remain visually identical, suggesting the numerical convergence of  VSC 2D spectra. When the CavMD propagation utilizes the fixed point-charge dipole model while the spectroscopic post-processing employs the DID dipole model, as shown in Figs. \ref{fig:2d_size_denpendence}g-i, the spectra remain largely unchanged under different system sizes. In all cases, analogous to Fig. \ref{fig:2d-ir-raman}c, the TIRV region becomes significantly distorted than that outside the cavity, suggesting again the importance of using the consistent dipole surface during both CavMD propagation and spectroscopic calculations.  Finally, the bottom two rows report the corresponding VSC 2D-IIR spectra at a larger effective coupling strength $\widetilde{\varepsilon}  \sqrt{N_{\rm{simu}}/64} = 7\times 10^{-4}$ a.u. Obviously, spectra convergence is reached under various molecular sizes.

\section{Conclusion}\label{sec:conclusion}

To summarize, we have extended the CavMD approach to study 2D-IIR spectra of liquid water under VSC by performing equilibrium-nonequilibrium CavMD simulations. Our study here reveals that to capture polariton spectra under VSC, the dipole moments employed in CavMD simulations should be consistent with the model used in spectroscopy simulations. In linear IR spectra, a mismatch between the dipole models in CavMD simulations and spectroscopic post-processing results in a  nonphysical, small middle peak between the two polariton branches. The effect is even more pronounced in 2D-IIR spectra, where such mismatch leads to strongly distorted and qualitatively wrong results. For instance, the TIRV region at the low-frequency IR domain, which is in principle unaffected by polariton formation at the high-frequency \ch{O-H} stretch band, is strongly distorted when different dipole surfaces are used. Therefore, both the CavMD simulations and spectroscopic post-processing should employ the same dipole surface.

Using the DID dipole model, we have further systematically compared the 2D-IIR spectra inside versus outside the cavity. It appears that, for liquid water, the cavity 2D-IIR spectra under VSC simply filters the outside-cavity spectra \cite{Schwennicke2024} by  splitting the molecular vibrational band coupled to the cavity to a pair of polariton branches, while significantly reducing the signals of other IR frequency regions. Although the 2D-IIR spectroscopy has not been utilized for studying VSC, this work provides the foundation for employing direct CavMD simulations for constructing experimentally more relevant 2D spectra under VSC in the near future.

 \section{Acknowledgments}
    This material is based upon work supported by the U.S. National Science Foundation under Grant No. CHE-2502758. This work used the Anvil HPC at Purdue University through allocation CHE250091 from the Advanced Cyberinfrastructure Coordination Ecosystem: Services \& Support (ACCESS) program, which is supported by U.S. National Science Foundation grants \#2138259, \#2138286, \#2138307, \#2137603, and \#2138296.

\section{Data Availability Statement}
    The source code for performing the simulations will be available at the main version of i-PI 3.0 package on Github upon the publication of this paper (pull request pending). Input files, simulation data, and post-processing scripts are archived in a different Github repository: \url{https://github.com/TaoELi/cavmd_examples_2dspectra}

\appendix

\section*{Appendix: The origin of the middle peak in linear polariton spectra}\label{sec:derivation}

For liquid water, according to Eq. \eqref{eq:mu_did},  the total dipole moment vector evaluated under the DID dipole moment, $\bm{\mu}(t)$, can be calculated by
\begin{equation}
    \bm{\mu}(t) = \sum_{nj} Q_{nj} \mathbf{R}_{nj}(t) + \sum_{njn'j'} \mathbf{R}_{n'j'}(t) \overset{\leftrightarrow}{\mathbf{D}}_{njn'j'} \mathbf{R}_{nj}(t)  ,
\end{equation}
where the indexes run over the $j=1,2,3$-th atom in molecule $n$. The first term on the right hand side comes from the contribution of each individual molecule, and the tensor $\overset{\leftrightarrow}{\mathbf{D}}_{njn'j'}$ accounts for the polarization effect induced by the neighboring molecules.
Denoting the time-dependent  collective variable $\mathbf{Z}_{nj}(t)$ as
\begin{equation}
    \mathbf{Z}_{nj}(t) = \sum_{n'j'} \mathbf{R}_{n'j'}(t) \frac{\overset{\leftrightarrow}{\mathbf{D}}_{njn'j'}}{Q_{nj}}  ,
\end{equation}
the above dipole vector becomes
\begin{equation}
        \bm{\mu}(t) = \sum_{nj} [Q_{nj} + \mathbf{Z}_{nj}(t)] \mathbf{R}_{nj}(t) .
\end{equation}
We now consider the $x$-component of the dipole vector, $\mu_x(t) = \bm{\mu}(t)\cdot \mathbf{e}_x$, which is assumed to interact with the cavity photon mode during CavMD simulations. Denoting $x_n(t)\equiv\sum_{j\in n}Q_{nj}\mathbf{R}_{nj}\cdot\mathbf{e}_x$ and $\bar{z}_n(t)\equiv \sum_{j\in n} \mathbf{Z}_{nj}(t) \mathbf{R}_{nj}\cdot \mathbf{e}_x / x_n(t)$, we may express $\mu_x(t)$ as
\begin{equation}\label{eq:mu_x_def}
    \mu_x(t) = \sum_{n} [1 + \bar{z}_n(t)] x_n(t) .
\end{equation}
If the fixed point-charge dipole model is used to propagate CavMD under VSC, $\sum_{n}x_n(t)$ defines the molecular bright-state coupled to the cavity; whereas all other linear combinations of $x_n(t)$, including $\sum_{n} \bar{z}_n(t) x_n(t) $, constitute the molecular dark states.

\begin{figure}[htbp]
\centering
\includegraphics[width=1.\linewidth]{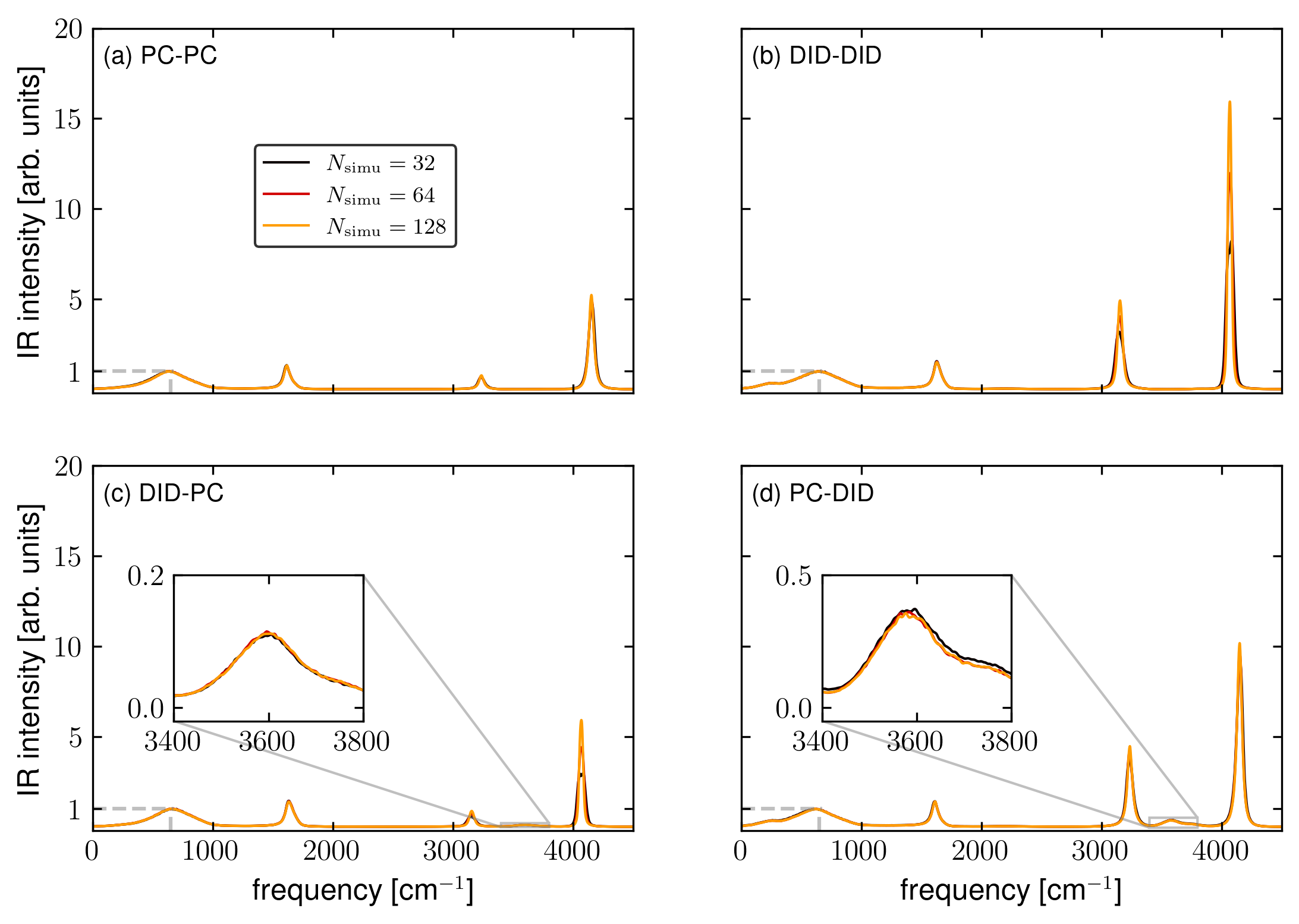}
\caption{\label{fig:1d_size_denpendence} \textbf{Simulated molecular IR spectra versus the system size, analogous to Fig. \ref{fig:1d-ir}.} In each part, the explicitly simulated molecular number is tuned from $N_{\rm{simu}}=$ 32 (black) to 64 (red) and 128 (orange). For fixing the Rabi splitting, the corresponding effective light-matter coupling strength is chosen as $\widetilde{\varepsilon} \cdot \sqrt{N_{\rm{simu}}/64}=7 \times 10^{-4}$ a.u.  All other simulation parameters remain the same as those in Fig. \ref{fig:1d-ir}. For facilitating the comparison between IR spectra, the water libration peaks are normalized to unity. 
}
\end{figure}

Apparently, $\mu_x(t)$, the dipole moment evaluated under the DID model, contains a small contribution from molecular dark states, if $\sum_{n}x_n(t)$  (or the point-charge dipole model) is used to couple to the cavity in simulations. This explains the emergence of the middle peak when the CavMD propagation and spectroscopic post-processing utilize different dipole models,  as shown in Figs. \ref{fig:1d-ir} c,d. 
Additionally, because the magnitude of $\bar{z}_n(t)$ (a local property of individual molecules) does not decay when the molecular number increases, this middle peak should also persist when the molecular system size increases, which is further verified in 
Fig. \ref{fig:1d_size_denpendence} in the Appendix.


\begin{thebibliography}{102}%
\makeatletter
\providecommand \@ifxundefined [1]{%
 \@ifx{#1\undefined}
}%
\providecommand \@ifnum [1]{%
 \ifnum #1\expandafter \@firstoftwo
 \else \expandafter \@secondoftwo
 \fi
}%
\providecommand \@ifx [1]{%
 \ifx #1\expandafter \@firstoftwo
 \else \expandafter \@secondoftwo
 \fi
}%
\providecommand \natexlab [1]{#1}%
\providecommand \enquote  [1]{``#1''}%
\providecommand \bibnamefont  [1]{#1}%
\providecommand \bibfnamefont [1]{#1}%
\providecommand \citenamefont [1]{#1}%
\providecommand \href@noop [0]{\@secondoftwo}%
\providecommand \href [0]{\begingroup \@sanitize@url \@href}%
\providecommand \@href[1]{\@@startlink{#1}\@@href}%
\providecommand \@@href[1]{\endgroup#1\@@endlink}%
\providecommand \@sanitize@url [0]{\catcode `\\12\catcode `\$12\catcode `\&12\catcode `\#12\catcode `\^12\catcode `\_12\catcode `\%12\relax}%
\providecommand \@@startlink[1]{}%
\providecommand \@@endlink[0]{}%
\providecommand \url  [0]{\begingroup\@sanitize@url \@url }%
\providecommand \@url [1]{\endgroup\@href {#1}{\urlprefix }}%
\providecommand \urlprefix  [0]{URL }%
\providecommand \Eprint [0]{\href }%
\providecommand \doibase [0]{http://dx.doi.org/}%
\providecommand \selectlanguage [0]{\@gobble}%
\providecommand \bibinfo  [0]{\@secondoftwo}%
\providecommand \bibfield  [0]{\@secondoftwo}%
\providecommand \translation [1]{[#1]}%
\providecommand \BibitemOpen [0]{}%
\providecommand \bibitemStop [0]{}%
\providecommand \bibitemNoStop [0]{.\EOS\space}%
\providecommand \EOS [0]{\spacefactor3000\relax}%
\providecommand \BibitemShut  [1]{\csname bibitem#1\endcsname}%
\let\auto@bib@innerbib\@empty
\bibitem [{\citenamefont {Shalabney}\ \emph {et~al.}(2015)\citenamefont {Shalabney}, \citenamefont {George}, \citenamefont {Hutchison}, \citenamefont {Pupillo}, \citenamefont {Genet},\ and\ \citenamefont {Ebbesen}}]{Shalabney2015}%
  \BibitemOpen
  \bibfield  {author} {\bibinfo {author} {\bibfnamefont {A.}~\bibnamefont {Shalabney}}, \bibinfo {author} {\bibfnamefont {J.}~\bibnamefont {George}}, \bibinfo {author} {\bibfnamefont {J.}~\bibnamefont {Hutchison}}, \bibinfo {author} {\bibfnamefont {G.}~\bibnamefont {Pupillo}}, \bibinfo {author} {\bibfnamefont {C.}~\bibnamefont {Genet}}, \ and\ \bibinfo {author} {\bibfnamefont {T.~W.}\ \bibnamefont {Ebbesen}},\ }\href {\doibase 10.1038/ncomms6981} {\bibfield  {journal} {\bibinfo  {journal} {Nat. Commun.}\ }\textbf {\bibinfo {volume} {6}},\ \bibinfo {pages} {5981} (\bibinfo {year} {2015})}\BibitemShut {NoStop}%
\bibitem [{\citenamefont {Long}\ and\ \citenamefont {Simpkins}(2015)}]{Long2015}%
  \BibitemOpen
  \bibfield  {author} {\bibinfo {author} {\bibfnamefont {J.~P.}\ \bibnamefont {Long}}\ and\ \bibinfo {author} {\bibfnamefont {B.~S.}\ \bibnamefont {Simpkins}},\ }\href {\doibase 10.1021/ph5003347} {\bibfield  {journal} {\bibinfo  {journal} {ACS Photonics}\ }\textbf {\bibinfo {volume} {2}},\ \bibinfo {pages} {130} (\bibinfo {year} {2015})}\BibitemShut {NoStop}%
\bibitem [{\citenamefont {Ebbesen}, \citenamefont {Rubio},\ and\ \citenamefont {Scholes}(2023)}]{Ebbesen2023}%
  \BibitemOpen
  \bibfield  {author} {\bibinfo {author} {\bibfnamefont {T.~W.}\ \bibnamefont {Ebbesen}}, \bibinfo {author} {\bibfnamefont {A.}~\bibnamefont {Rubio}}, \ and\ \bibinfo {author} {\bibfnamefont {G.~D.}\ \bibnamefont {Scholes}},\ }\href {\doibase 10.1021/ACS.CHEMREV.3C00637} {\bibfield  {journal} {\bibinfo  {journal} {Chem. Rev.}\ }\textbf {\bibinfo {volume} {123}},\ \bibinfo {pages} {12037} (\bibinfo {year} {2023})}\BibitemShut {NoStop}%
\bibitem [{\citenamefont {Wright}, \citenamefont {Nelson},\ and\ \citenamefont {Weichman}(2023)}]{Wright2023}%
  \BibitemOpen
  \bibfield  {author} {\bibinfo {author} {\bibfnamefont {A.~D.}\ \bibnamefont {Wright}}, \bibinfo {author} {\bibfnamefont {J.~C.}\ \bibnamefont {Nelson}}, \ and\ \bibinfo {author} {\bibfnamefont {M.~L.}\ \bibnamefont {Weichman}},\ }\href {\doibase 10.1021/jacs.3c00126} {\bibfield  {journal} {\bibinfo  {journal} {J. Am. Chem. Soc.}\ }\textbf {\bibinfo {volume} {145}},\ \bibinfo {pages} {5982} (\bibinfo {year} {2023})}\BibitemShut {NoStop}%
\bibitem [{\citenamefont {Thomas}\ \emph {et~al.}(2016)\citenamefont {Thomas}, \citenamefont {George}, \citenamefont {Shalabney}, \citenamefont {Dryzhakov}, \citenamefont {Varma}, \citenamefont {Moran}, \citenamefont {Chervy}, \citenamefont {Zhong}, \citenamefont {Devaux}, \citenamefont {Genet}, \citenamefont {Hutchison},\ and\ \citenamefont {Ebbesen}}]{thomasGroundStateChemicalReactivity2016}%
  \BibitemOpen
  \bibfield  {author} {\bibinfo {author} {\bibfnamefont {A.}~\bibnamefont {Thomas}}, \bibinfo {author} {\bibfnamefont {J.}~\bibnamefont {George}}, \bibinfo {author} {\bibfnamefont {A.}~\bibnamefont {Shalabney}}, \bibinfo {author} {\bibfnamefont {M.}~\bibnamefont {Dryzhakov}}, \bibinfo {author} {\bibfnamefont {S.~J.}\ \bibnamefont {Varma}}, \bibinfo {author} {\bibfnamefont {J.}~\bibnamefont {Moran}}, \bibinfo {author} {\bibfnamefont {T.}~\bibnamefont {Chervy}}, \bibinfo {author} {\bibfnamefont {X.}~\bibnamefont {Zhong}}, \bibinfo {author} {\bibfnamefont {E.}~\bibnamefont {Devaux}}, \bibinfo {author} {\bibfnamefont {C.}~\bibnamefont {Genet}}, \bibinfo {author} {\bibfnamefont {J.~A.}\ \bibnamefont {Hutchison}}, \ and\ \bibinfo {author} {\bibfnamefont {T.~W.}\ \bibnamefont {Ebbesen}},\ }\href {\doibase 10.1002/anie.201605504} {\bibfield  {journal} {\bibinfo  {journal} {Angew. Chem. Int. Ed.}\ }\textbf {\bibinfo {volume} {55}},\ \bibinfo {pages} {11462} (\bibinfo {year} {2016})}\BibitemShut {NoStop}%
\bibitem [{\citenamefont {Thomas}\ \emph {et~al.}(2019)\citenamefont {Thomas}, \citenamefont {Lethuillier-Karl}, \citenamefont {Nagarajan}, \citenamefont {Vergauwe}, \citenamefont {George}, \citenamefont {Chervy}, \citenamefont {Shalabney}, \citenamefont {Devaux}, \citenamefont {Genet}, \citenamefont {Moran},\ and\ \citenamefont {Ebbesen}}]{Thomas2019_science}%
  \BibitemOpen
  \bibfield  {author} {\bibinfo {author} {\bibfnamefont {A.}~\bibnamefont {Thomas}}, \bibinfo {author} {\bibfnamefont {L.}~\bibnamefont {Lethuillier-Karl}}, \bibinfo {author} {\bibfnamefont {K.}~\bibnamefont {Nagarajan}}, \bibinfo {author} {\bibfnamefont {R.~M.~A.}\ \bibnamefont {Vergauwe}}, \bibinfo {author} {\bibfnamefont {J.}~\bibnamefont {George}}, \bibinfo {author} {\bibfnamefont {T.}~\bibnamefont {Chervy}}, \bibinfo {author} {\bibfnamefont {A.}~\bibnamefont {Shalabney}}, \bibinfo {author} {\bibfnamefont {E.}~\bibnamefont {Devaux}}, \bibinfo {author} {\bibfnamefont {C.}~\bibnamefont {Genet}}, \bibinfo {author} {\bibfnamefont {J.}~\bibnamefont {Moran}}, \ and\ \bibinfo {author} {\bibfnamefont {T.~W.}\ \bibnamefont {Ebbesen}},\ }\href {\doibase 10.1126/science.aau7742} {\bibfield  {journal} {\bibinfo  {journal} {Science}\ }\textbf {\bibinfo {volume} {363}},\ \bibinfo {pages} {615} (\bibinfo {year} {2019})}\BibitemShut {NoStop}%
\bibitem [{\citenamefont {Ahn}\ \emph {et~al.}(2023)\citenamefont {Ahn}, \citenamefont {Triana}, \citenamefont {Recabal}, \citenamefont {Herrera},\ and\ \citenamefont {Simpkins}}]{Ahn2023Science}%
  \BibitemOpen
  \bibfield  {author} {\bibinfo {author} {\bibfnamefont {W.}~\bibnamefont {Ahn}}, \bibinfo {author} {\bibfnamefont {J.~F.}\ \bibnamefont {Triana}}, \bibinfo {author} {\bibfnamefont {F.}~\bibnamefont {Recabal}}, \bibinfo {author} {\bibfnamefont {F.}~\bibnamefont {Herrera}}, \ and\ \bibinfo {author} {\bibfnamefont {B.~S.}\ \bibnamefont {Simpkins}},\ }\href {\doibase 10.1126/science.ade7147} {\bibfield  {journal} {\bibinfo  {journal} {Science}\ }\textbf {\bibinfo {volume} {380}},\ \bibinfo {pages} {1165} (\bibinfo {year} {2023})}\BibitemShut {NoStop}%
\bibitem [{\citenamefont {Sandeep}\ \emph {et~al.}(2025)\citenamefont {Sandeep}, \citenamefont {Swaminathan}, \citenamefont {Jayachandran}, \citenamefont {Nagarajan}, \citenamefont {Gautier}, \citenamefont {Kushida}, \citenamefont {Chervy}, \citenamefont {Vergauwe}, \citenamefont {Thomas},\ and\ \citenamefont {Ebbesen}}]{sandeepClusterFormationPhase}%
  \BibitemOpen
  \bibfield  {author} {\bibinfo {author} {\bibfnamefont {K.}~\bibnamefont {Sandeep}}, \bibinfo {author} {\bibfnamefont {S.}~\bibnamefont {Swaminathan}}, \bibinfo {author} {\bibfnamefont {A.}~\bibnamefont {Jayachandran}}, \bibinfo {author} {\bibfnamefont {K.}~\bibnamefont {Nagarajan}}, \bibinfo {author} {\bibfnamefont {J.}~\bibnamefont {Gautier}}, \bibinfo {author} {\bibfnamefont {S.}~\bibnamefont {Kushida}}, \bibinfo {author} {\bibfnamefont {T.}~\bibnamefont {Chervy}}, \bibinfo {author} {\bibfnamefont {R.}~\bibnamefont {Vergauwe}}, \bibinfo {author} {\bibfnamefont {A.}~\bibnamefont {Thomas}}, \ and\ \bibinfo {author} {\bibfnamefont {T.}~\bibnamefont {Ebbesen}},\ }\href {\doibase 10.1002/ange.202516917} {\bibfield  {journal} {\bibinfo  {journal} {Angew. Chem.}\ }\textbf {\bibinfo {volume} {2025}},\ \bibinfo {pages} {e16917} (\bibinfo {year} {2025})}\BibitemShut {NoStop}%
\bibitem [{\citenamefont {Hirai}\ \emph {et~al.}(2021)\citenamefont {Hirai}, \citenamefont {Ishikawa}, \citenamefont {Chervy}, \citenamefont {Hutchison},\ and\ \citenamefont {Uji-i}}]{hirai2021selective}%
  \BibitemOpen
  \bibfield  {author} {\bibinfo {author} {\bibfnamefont {K.}~\bibnamefont {Hirai}}, \bibinfo {author} {\bibfnamefont {H.}~\bibnamefont {Ishikawa}}, \bibinfo {author} {\bibfnamefont {T.}~\bibnamefont {Chervy}}, \bibinfo {author} {\bibfnamefont {J.~A.}\ \bibnamefont {Hutchison}}, \ and\ \bibinfo {author} {\bibfnamefont {H.}~\bibnamefont {Uji-i}},\ }\href {\doibase 10.1039/D1SC03706D} {\bibfield  {journal} {\bibinfo  {journal} {Chem. Sci.}\ }\textbf {\bibinfo {volume} {12}},\ \bibinfo {pages} {11986} (\bibinfo {year} {2021})}\BibitemShut {NoStop}%
\bibitem [{\citenamefont {Sandeep}\ \emph {et~al.}(2022)\citenamefont {Sandeep}, \citenamefont {Joseph}, \citenamefont {Gautier}, \citenamefont {Nagarajan}, \citenamefont {Sujith}, \citenamefont {Thomas},\ and\ \citenamefont {Ebbesen}}]{sandeepManipulatingSelfAssemblyPhenyleneethynylenes2022}%
  \BibitemOpen
  \bibfield  {author} {\bibinfo {author} {\bibfnamefont {K.}~\bibnamefont {Sandeep}}, \bibinfo {author} {\bibfnamefont {K.}~\bibnamefont {Joseph}}, \bibinfo {author} {\bibfnamefont {J.}~\bibnamefont {Gautier}}, \bibinfo {author} {\bibfnamefont {K.}~\bibnamefont {Nagarajan}}, \bibinfo {author} {\bibfnamefont {M.}~\bibnamefont {Sujith}}, \bibinfo {author} {\bibfnamefont {K.~G.}\ \bibnamefont {Thomas}}, \ and\ \bibinfo {author} {\bibfnamefont {T.~W.}\ \bibnamefont {Ebbesen}},\ }\href {\doibase 10.1021/acs.jpclett.1c03893} {\bibfield  {journal} {\bibinfo  {journal} {J. Phys. Chem. Lett.}\ }\textbf {\bibinfo {volume} {13}},\ \bibinfo {pages} {1209} (\bibinfo {year} {2022})}\BibitemShut {NoStop}%
\bibitem [{\citenamefont {Joseph}\ \emph {et~al.}(2021)\citenamefont {Joseph}, \citenamefont {Kushida}, \citenamefont {Smarsly}, \citenamefont {Ihiawakrim}, \citenamefont {Thomas}, \citenamefont {{Paravicini-Bagliani}}, \citenamefont {Nagarajan}, \citenamefont {Vergauwe}, \citenamefont {Devaux}, \citenamefont {Ersen}, \citenamefont {Bunz},\ and\ \citenamefont {Ebbesen}}]{josephSupramolecularAssemblyConjugated2021}%
  \BibitemOpen
  \bibfield  {author} {\bibinfo {author} {\bibfnamefont {K.}~\bibnamefont {Joseph}}, \bibinfo {author} {\bibfnamefont {S.}~\bibnamefont {Kushida}}, \bibinfo {author} {\bibfnamefont {E.}~\bibnamefont {Smarsly}}, \bibinfo {author} {\bibfnamefont {D.}~\bibnamefont {Ihiawakrim}}, \bibinfo {author} {\bibfnamefont {A.}~\bibnamefont {Thomas}}, \bibinfo {author} {\bibfnamefont {G.~L.}\ \bibnamefont {{Paravicini-Bagliani}}}, \bibinfo {author} {\bibfnamefont {K.}~\bibnamefont {Nagarajan}}, \bibinfo {author} {\bibfnamefont {R.}~\bibnamefont {Vergauwe}}, \bibinfo {author} {\bibfnamefont {E.}~\bibnamefont {Devaux}}, \bibinfo {author} {\bibfnamefont {O.}~\bibnamefont {Ersen}}, \bibinfo {author} {\bibfnamefont {U.~H.~F.}\ \bibnamefont {Bunz}}, \ and\ \bibinfo {author} {\bibfnamefont {T.~W.}\ \bibnamefont {Ebbesen}},\ }\href {\doibase 10.1002/anie.202105840} {\bibfield  {journal} {\bibinfo  {journal} {Angew. Chem. Int. Ed.}\ }\textbf {\bibinfo {volume} {60}},\ \bibinfo {pages} {19665} (\bibinfo {year} {2021})}\BibitemShut
  {NoStop}%
\bibitem [{\citenamefont {Xiang}\ \emph {et~al.}(2019)\citenamefont {Xiang}, \citenamefont {Ribeiro}, \citenamefont {Chen}, \citenamefont {Wang}, \citenamefont {Du}, \citenamefont {{Yuen-Zhou}},\ and\ \citenamefont {Xiong}}]{xiangStateSelectivePolaritonDark2019}%
  \BibitemOpen
  \bibfield  {author} {\bibinfo {author} {\bibfnamefont {B.}~\bibnamefont {Xiang}}, \bibinfo {author} {\bibfnamefont {R.~F.}\ \bibnamefont {Ribeiro}}, \bibinfo {author} {\bibfnamefont {L.}~\bibnamefont {Chen}}, \bibinfo {author} {\bibfnamefont {J.}~\bibnamefont {Wang}}, \bibinfo {author} {\bibfnamefont {M.}~\bibnamefont {Du}}, \bibinfo {author} {\bibfnamefont {J.}~\bibnamefont {{Yuen-Zhou}}}, \ and\ \bibinfo {author} {\bibfnamefont {W.}~\bibnamefont {Xiong}},\ }\href {\doibase 10.1021/acs.jpca.9b04601} {\bibfield  {journal} {\bibinfo  {journal} {J. Phys. Chem. A}\ }\textbf {\bibinfo {volume} {123}},\ \bibinfo {pages} {5918} (\bibinfo {year} {2019})}\BibitemShut {NoStop}%
\bibitem [{\citenamefont {Xiang}\ \emph {et~al.}(2020)\citenamefont {Xiang}, \citenamefont {Ribeiro}, \citenamefont {Du}, \citenamefont {Chen}, \citenamefont {Yang}, \citenamefont {Wang}, \citenamefont {Yuen-Zhou},\ and\ \citenamefont {Xiong}}]{Xiang2020Science}%
  \BibitemOpen
  \bibfield  {author} {\bibinfo {author} {\bibfnamefont {B.}~\bibnamefont {Xiang}}, \bibinfo {author} {\bibfnamefont {R.~F.}\ \bibnamefont {Ribeiro}}, \bibinfo {author} {\bibfnamefont {M.}~\bibnamefont {Du}}, \bibinfo {author} {\bibfnamefont {L.}~\bibnamefont {Chen}}, \bibinfo {author} {\bibfnamefont {Z.}~\bibnamefont {Yang}}, \bibinfo {author} {\bibfnamefont {J.}~\bibnamefont {Wang}}, \bibinfo {author} {\bibfnamefont {J.}~\bibnamefont {Yuen-Zhou}}, \ and\ \bibinfo {author} {\bibfnamefont {W.}~\bibnamefont {Xiong}},\ }\href {\doibase 10.1126/science.aba3544} {\bibfield  {journal} {\bibinfo  {journal} {Science}\ }\textbf {\bibinfo {volume} {368}},\ \bibinfo {pages} {665} (\bibinfo {year} {2020})}\BibitemShut {NoStop}%
\bibitem [{\citenamefont {Grafton}\ \emph {et~al.}(2021)\citenamefont {Grafton}, \citenamefont {Dunkelberger}, \citenamefont {Simpkins}, \citenamefont {Triana}, \citenamefont {Hern{\'{a}}ndez}, \citenamefont {Herrera},\ and\ \citenamefont {Owrutsky}}]{Grafton2020}%
  \BibitemOpen
  \bibfield  {author} {\bibinfo {author} {\bibfnamefont {A.~B.}\ \bibnamefont {Grafton}}, \bibinfo {author} {\bibfnamefont {A.~D.}\ \bibnamefont {Dunkelberger}}, \bibinfo {author} {\bibfnamefont {B.~S.}\ \bibnamefont {Simpkins}}, \bibinfo {author} {\bibfnamefont {J.~F.}\ \bibnamefont {Triana}}, \bibinfo {author} {\bibfnamefont {F.~J.}\ \bibnamefont {Hern{\'{a}}ndez}}, \bibinfo {author} {\bibfnamefont {F.}~\bibnamefont {Herrera}}, \ and\ \bibinfo {author} {\bibfnamefont {J.~C.}\ \bibnamefont {Owrutsky}},\ }\href {\doibase 10.1038/s41467-020-20535-z} {\bibfield  {journal} {\bibinfo  {journal} {Nat. Commun.}\ }\textbf {\bibinfo {volume} {12}},\ \bibinfo {pages} {214} (\bibinfo {year} {2021})}\BibitemShut {NoStop}%
\bibitem [{\citenamefont {Xiong}(2023)}]{xiongMolecularVibrationalPolariton2023}%
  \BibitemOpen
  \bibfield  {author} {\bibinfo {author} {\bibfnamefont {W.}~\bibnamefont {Xiong}},\ }\href {\doibase 10.1021/acs.accounts.2c00796} {\bibfield  {journal} {\bibinfo  {journal} {Acc. Chem. Res.}\ }\textbf {\bibinfo {volume} {56}},\ \bibinfo {pages} {776} (\bibinfo {year} {2023})}\BibitemShut {NoStop}%
\bibitem [{\citenamefont {Simpkins}, \citenamefont {Dunkelberger},\ and\ \citenamefont {Vurgaftman}(2023)}]{Simpkins2023}%
  \BibitemOpen
  \bibfield  {author} {\bibinfo {author} {\bibfnamefont {B.~S.}\ \bibnamefont {Simpkins}}, \bibinfo {author} {\bibfnamefont {A.~D.}\ \bibnamefont {Dunkelberger}}, \ and\ \bibinfo {author} {\bibfnamefont {I.}~\bibnamefont {Vurgaftman}},\ }\href {\doibase 10.1021/acs.chemrev.2c00774} {\bibfield  {journal} {\bibinfo  {journal} {Chem. Rev.}\ }\textbf {\bibinfo {volume} {123}},\ \bibinfo {pages} {5020} (\bibinfo {year} {2023})}\BibitemShut {NoStop}%
\bibitem [{\citenamefont {Xiang}\ and\ \citenamefont {Xiong}(2024)}]{Xiang2024}%
  \BibitemOpen
  \bibfield  {author} {\bibinfo {author} {\bibfnamefont {B.}~\bibnamefont {Xiang}}\ and\ \bibinfo {author} {\bibfnamefont {W.}~\bibnamefont {Xiong}},\ }\href {\doibase 10.1021/ACS.CHEMREV.3C00662/ASSET/IMAGES/LARGE/CR3C00662_0036.JPEG} {\bibfield  {journal} {\bibinfo  {journal} {Chem. Rev.}\ }\textbf {\bibinfo {volume} {124}},\ \bibinfo {pages} {2512} (\bibinfo {year} {2024})}\BibitemShut {NoStop}%
\bibitem [{\citenamefont {Chen}\ \emph {et~al.}(2022)\citenamefont {Chen}, \citenamefont {Du}, \citenamefont {Yang}, \citenamefont {{Yuen-Zhou}},\ and\ \citenamefont {Xiong}}]{chenCavityenabledEnhancementUltrafast2022}%
  \BibitemOpen
  \bibfield  {author} {\bibinfo {author} {\bibfnamefont {T.-T.}\ \bibnamefont {Chen}}, \bibinfo {author} {\bibfnamefont {M.}~\bibnamefont {Du}}, \bibinfo {author} {\bibfnamefont {Z.}~\bibnamefont {Yang}}, \bibinfo {author} {\bibfnamefont {J.}~\bibnamefont {{Yuen-Zhou}}}, \ and\ \bibinfo {author} {\bibfnamefont {W.}~\bibnamefont {Xiong}},\ }\href {\doibase 10.1126/science.add0276} {\bibfield  {journal} {\bibinfo  {journal} {Science}\ }\textbf {\bibinfo {volume} {378}},\ \bibinfo {pages} {790} (\bibinfo {year} {2022})}\BibitemShut {NoStop}%
\bibitem [{\citenamefont {Yin}\ \emph {et~al.}(2025)\citenamefont {Yin}, \citenamefont {Liu}, \citenamefont {Zhang}, \citenamefont {Sheng}, \citenamefont {Mao},\ and\ \citenamefont {Xiong}}]{yinOvercomingEnergyDisorder2025}%
  \BibitemOpen
  \bibfield  {author} {\bibinfo {author} {\bibfnamefont {G.}~\bibnamefont {Yin}}, \bibinfo {author} {\bibfnamefont {T.}~\bibnamefont {Liu}}, \bibinfo {author} {\bibfnamefont {L.}~\bibnamefont {Zhang}}, \bibinfo {author} {\bibfnamefont {T.}~\bibnamefont {Sheng}}, \bibinfo {author} {\bibfnamefont {H.}~\bibnamefont {Mao}}, \ and\ \bibinfo {author} {\bibfnamefont {W.}~\bibnamefont {Xiong}},\ }\href {\doibase 10.1126/science.adx3137} {\bibfield  {journal} {\bibinfo  {journal} {Science}\ }\textbf {\bibinfo {volume} {389}},\ \bibinfo {pages} {845} (\bibinfo {year} {2025})}\BibitemShut {NoStop}%
\bibitem [{\citenamefont {Dunkelberger}\ \emph {et~al.}(2016)\citenamefont {Dunkelberger}, \citenamefont {Spann}, \citenamefont {Fears}, \citenamefont {Simpkins},\ and\ \citenamefont {Owrutsky}}]{Dunkelberger2016}%
  \BibitemOpen
  \bibfield  {author} {\bibinfo {author} {\bibfnamefont {A.~D.}\ \bibnamefont {Dunkelberger}}, \bibinfo {author} {\bibfnamefont {B.~T.}\ \bibnamefont {Spann}}, \bibinfo {author} {\bibfnamefont {K.~P.}\ \bibnamefont {Fears}}, \bibinfo {author} {\bibfnamefont {B.~S.}\ \bibnamefont {Simpkins}}, \ and\ \bibinfo {author} {\bibfnamefont {J.~C.}\ \bibnamefont {Owrutsky}},\ }\href {\doibase 10.1038/ncomms13504} {\bibfield  {journal} {\bibinfo  {journal} {Nat. Commun.}\ }\textbf {\bibinfo {volume} {7}},\ \bibinfo {pages} {1} (\bibinfo {year} {2016})}\BibitemShut {NoStop}%
\bibitem [{\citenamefont {Pyles}\ \emph {et~al.}(2024)\citenamefont {Pyles}, \citenamefont {Simpkins}, \citenamefont {Vurgaftman}, \citenamefont {Owrutsky},\ and\ \citenamefont {Dunkelberger}}]{Pyles2024}%
  \BibitemOpen
  \bibfield  {author} {\bibinfo {author} {\bibfnamefont {C.~G.}\ \bibnamefont {Pyles}}, \bibinfo {author} {\bibfnamefont {B.~S.}\ \bibnamefont {Simpkins}}, \bibinfo {author} {\bibfnamefont {I.}~\bibnamefont {Vurgaftman}}, \bibinfo {author} {\bibfnamefont {J.~C.}\ \bibnamefont {Owrutsky}}, \ and\ \bibinfo {author} {\bibfnamefont {A.~D.}\ \bibnamefont {Dunkelberger}},\ }\href {\doibase 10.1063/5.0239301/3326988} {\bibfield  {journal} {\bibinfo  {journal} {J. Chem. Phys.}\ }\textbf {\bibinfo {volume} {161}},\ \bibinfo {pages} {234202} (\bibinfo {year} {2024})}\BibitemShut {NoStop}%
\bibitem [{\citenamefont {Luk}\ \emph {et~al.}(2017)\citenamefont {Luk}, \citenamefont {Feist}, \citenamefont {Toppari},\ and\ \citenamefont {Groenhof}}]{Luk2017}%
  \BibitemOpen
  \bibfield  {author} {\bibinfo {author} {\bibfnamefont {H.~L.}\ \bibnamefont {Luk}}, \bibinfo {author} {\bibfnamefont {J.}~\bibnamefont {Feist}}, \bibinfo {author} {\bibfnamefont {J.~J.}\ \bibnamefont {Toppari}}, \ and\ \bibinfo {author} {\bibfnamefont {G.}~\bibnamefont {Groenhof}},\ }\href {\doibase 10.1021/acs.jctc.7b00388} {\bibfield  {journal} {\bibinfo  {journal} {J. Chem. Theory Comput.}\ }\textbf {\bibinfo {volume} {13}},\ \bibinfo {pages} {4324} (\bibinfo {year} {2017})}\BibitemShut {NoStop}%
\bibitem [{\citenamefont {Flick}\ \emph {et~al.}(2017)\citenamefont {Flick}, \citenamefont {Ruggenthaler}, \citenamefont {Appel},\ and\ \citenamefont {Rubio}}]{flickAtomsMoleculesCavities2017}%
  \BibitemOpen
  \bibfield  {author} {\bibinfo {author} {\bibfnamefont {J.}~\bibnamefont {Flick}}, \bibinfo {author} {\bibfnamefont {M.}~\bibnamefont {Ruggenthaler}}, \bibinfo {author} {\bibfnamefont {H.}~\bibnamefont {Appel}}, \ and\ \bibinfo {author} {\bibfnamefont {A.}~\bibnamefont {Rubio}},\ }\href {\doibase 10.1073/pnas.1615509114} {\bibfield  {journal} {\bibinfo  {journal} {Proc. Natl. Acad. Sci. U.S.A.}\ }\textbf {\bibinfo {volume} {114}},\ \bibinfo {pages} {3026} (\bibinfo {year} {2017})}\BibitemShut {NoStop}%
\bibitem [{\citenamefont {Hoffmann}\ \emph {et~al.}(2018)\citenamefont {Hoffmann}, \citenamefont {Appel}, \citenamefont {Rubio},\ and\ \citenamefont {Maitra}}]{Hoffmann2018}%
  \BibitemOpen
  \bibfield  {author} {\bibinfo {author} {\bibfnamefont {N.~M.}\ \bibnamefont {Hoffmann}}, \bibinfo {author} {\bibfnamefont {H.}~\bibnamefont {Appel}}, \bibinfo {author} {\bibfnamefont {A.}~\bibnamefont {Rubio}}, \ and\ \bibinfo {author} {\bibfnamefont {N.~T.}\ \bibnamefont {Maitra}},\ }\href {\doibase 10.1140/epjb/e2018-90177-6} {\bibfield  {journal} {\bibinfo  {journal} {Eur. Phys. J. B}\ }\textbf {\bibinfo {volume} {91}},\ \bibinfo {pages} {180} (\bibinfo {year} {2018})}\BibitemShut {NoStop}%
\bibitem [{\citenamefont {Triana}, \citenamefont {Hern{\'a}ndez},\ and\ \citenamefont {Herrera}(2020)}]{trianaShapeElectricDipole2020}%
  \BibitemOpen
  \bibfield  {author} {\bibinfo {author} {\bibfnamefont {J.~F.}\ \bibnamefont {Triana}}, \bibinfo {author} {\bibfnamefont {F.~J.}\ \bibnamefont {Hern{\'a}ndez}}, \ and\ \bibinfo {author} {\bibfnamefont {F.}~\bibnamefont {Herrera}},\ }\href {\doibase 10.1063/5.0009869} {\bibfield  {journal} {\bibinfo  {journal} {J. Chem. Phys.}\ }\textbf {\bibinfo {volume} {152}},\ \bibinfo {pages} {234111} (\bibinfo {year} {2020})}\BibitemShut {NoStop}%
\bibitem [{\citenamefont {Botzung}\ \emph {et~al.}(2020)\citenamefont {Botzung}, \citenamefont {Hagenm{\"{u}}ller}, \citenamefont {Sch{\"{u}}tz}, \citenamefont {Dubail}, \citenamefont {Pupillo},\ and\ \citenamefont {Schachenmayer}}]{Botzung2020}%
  \BibitemOpen
  \bibfield  {author} {\bibinfo {author} {\bibfnamefont {T.}~\bibnamefont {Botzung}}, \bibinfo {author} {\bibfnamefont {D.}~\bibnamefont {Hagenm{\"{u}}ller}}, \bibinfo {author} {\bibfnamefont {S.}~\bibnamefont {Sch{\"{u}}tz}}, \bibinfo {author} {\bibfnamefont {J.}~\bibnamefont {Dubail}}, \bibinfo {author} {\bibfnamefont {G.}~\bibnamefont {Pupillo}}, \ and\ \bibinfo {author} {\bibfnamefont {J.}~\bibnamefont {Schachenmayer}},\ }\href {\doibase 10.1103/PhysRevB.102.144202} {\bibfield  {journal} {\bibinfo  {journal} {Phys. Rev. B}\ }\textbf {\bibinfo {volume} {102}},\ \bibinfo {pages} {144202} (\bibinfo {year} {2020})}\BibitemShut {NoStop}%
\bibitem [{\citenamefont {Fregoni}, \citenamefont {Garcia-Vidal},\ and\ \citenamefont {Feist}(2022)}]{Fregoni2022}%
  \BibitemOpen
  \bibfield  {author} {\bibinfo {author} {\bibfnamefont {J.}~\bibnamefont {Fregoni}}, \bibinfo {author} {\bibfnamefont {F.~J.}\ \bibnamefont {Garcia-Vidal}}, \ and\ \bibinfo {author} {\bibfnamefont {J.}~\bibnamefont {Feist}},\ }\href {\doibase 10.1021/acsphotonics.1c01749} {\bibfield  {journal} {\bibinfo  {journal} {ACS Photonics}\ }\textbf {\bibinfo {volume} {9}},\ \bibinfo {pages} {1096} (\bibinfo {year} {2022})}\BibitemShut {NoStop}%
\bibitem [{\citenamefont {Mandal}\ \emph {et~al.}(2023)\citenamefont {Mandal}, \citenamefont {Taylor}, \citenamefont {Weight}, \citenamefont {Koessler}, \citenamefont {Li},\ and\ \citenamefont {Huo}}]{Mandal2023ChemRev}%
  \BibitemOpen
  \bibfield  {author} {\bibinfo {author} {\bibfnamefont {A.}~\bibnamefont {Mandal}}, \bibinfo {author} {\bibfnamefont {M.~A.}\ \bibnamefont {Taylor}}, \bibinfo {author} {\bibfnamefont {B.~M.}\ \bibnamefont {Weight}}, \bibinfo {author} {\bibfnamefont {E.~R.}\ \bibnamefont {Koessler}}, \bibinfo {author} {\bibfnamefont {X.}~\bibnamefont {Li}}, \ and\ \bibinfo {author} {\bibfnamefont {P.}~\bibnamefont {Huo}},\ }\href {\doibase 10.1021/acs.chemrev.2c00855} {\bibfield  {journal} {\bibinfo  {journal} {Chem. Rev.}\ }\textbf {\bibinfo {volume} {123}},\ \bibinfo {pages} {9786} (\bibinfo {year} {2023})}\BibitemShut {NoStop}%
\bibitem [{\citenamefont {Yang}\ and\ \citenamefont {Cao}(2021)}]{YangCao2021}%
  \BibitemOpen
  \bibfield  {author} {\bibinfo {author} {\bibfnamefont {P.~Y.}\ \bibnamefont {Yang}}\ and\ \bibinfo {author} {\bibfnamefont {J.}~\bibnamefont {Cao}},\ }\href {\doibase 10.1021/ACS.JPCLETT.1C02210/SUPPL_FILE/JZ1C02210_SI_001.PDF} {\bibfield  {journal} {\bibinfo  {journal} {J. Phys. Chem. Lett.}\ }\textbf {\bibinfo {volume} {12}},\ \bibinfo {pages} {9531} (\bibinfo {year} {2021})}\BibitemShut {NoStop}%
\bibitem [{\citenamefont {Yang}\ \emph {et~al.}(2021)\citenamefont {Yang}, \citenamefont {Ou}, \citenamefont {Pei}, \citenamefont {Wang}, \citenamefont {Weng}, \citenamefont {Shuai}, \citenamefont {Mullen},\ and\ \citenamefont {Shao}}]{Yang2021QEDFT}%
  \BibitemOpen
  \bibfield  {author} {\bibinfo {author} {\bibfnamefont {J.}~\bibnamefont {Yang}}, \bibinfo {author} {\bibfnamefont {Q.}~\bibnamefont {Ou}}, \bibinfo {author} {\bibfnamefont {Z.}~\bibnamefont {Pei}}, \bibinfo {author} {\bibfnamefont {H.}~\bibnamefont {Wang}}, \bibinfo {author} {\bibfnamefont {B.}~\bibnamefont {Weng}}, \bibinfo {author} {\bibfnamefont {Z.}~\bibnamefont {Shuai}}, \bibinfo {author} {\bibfnamefont {K.}~\bibnamefont {Mullen}}, \ and\ \bibinfo {author} {\bibfnamefont {Y.}~\bibnamefont {Shao}},\ }\href {\doibase 10.1063/5.0057542} {\bibfield  {journal} {\bibinfo  {journal} {J. Chem. Phys.}\ }\textbf {\bibinfo {volume} {155}},\ \bibinfo {pages} {064107} (\bibinfo {year} {2021})}\BibitemShut {NoStop}%
\bibitem [{\citenamefont {Sokolovskii}\ \emph {et~al.}(2023)\citenamefont {Sokolovskii}, \citenamefont {Tichauer}, \citenamefont {Morozov}, \citenamefont {Feist},\ and\ \citenamefont {Groenhof}}]{Sokolovskii2022tmp}%
  \BibitemOpen
  \bibfield  {author} {\bibinfo {author} {\bibfnamefont {I.}~\bibnamefont {Sokolovskii}}, \bibinfo {author} {\bibfnamefont {R.~H.}\ \bibnamefont {Tichauer}}, \bibinfo {author} {\bibfnamefont {D.}~\bibnamefont {Morozov}}, \bibinfo {author} {\bibfnamefont {J.}~\bibnamefont {Feist}}, \ and\ \bibinfo {author} {\bibfnamefont {G.}~\bibnamefont {Groenhof}},\ }\href {\doibase 10.1038/s41467-023-42067-y} {\bibfield  {journal} {\bibinfo  {journal} {Nat. Commun.}\ }\textbf {\bibinfo {volume} {14}},\ \bibinfo {pages} {6613} (\bibinfo {year} {2023})}\BibitemShut {NoStop}%
\bibitem [{\citenamefont {Riso}\ \emph {et~al.}(2022)\citenamefont {Riso}, \citenamefont {Haugland}, \citenamefont {Ronca},\ and\ \citenamefont {Koch}}]{Riso2022}%
  \BibitemOpen
  \bibfield  {author} {\bibinfo {author} {\bibfnamefont {R.~R.}\ \bibnamefont {Riso}}, \bibinfo {author} {\bibfnamefont {T.~S.}\ \bibnamefont {Haugland}}, \bibinfo {author} {\bibfnamefont {E.}~\bibnamefont {Ronca}}, \ and\ \bibinfo {author} {\bibfnamefont {H.}~\bibnamefont {Koch}},\ }\href {\doibase 10.1038/s41467-022-29003-2} {\bibfield  {journal} {\bibinfo  {journal} {Nat. Commun.}\ }\textbf {\bibinfo {volume} {13}},\ \bibinfo {pages} {1368} (\bibinfo {year} {2022})}\BibitemShut {NoStop}%
\bibitem [{\citenamefont {Suyabatmaz}\ and\ \citenamefont {Ribeiro}(2023)}]{Suyabatmaz2023}%
  \BibitemOpen
  \bibfield  {author} {\bibinfo {author} {\bibfnamefont {E.}~\bibnamefont {Suyabatmaz}}\ and\ \bibinfo {author} {\bibfnamefont {R.~F.}\ \bibnamefont {Ribeiro}},\ }\href {\doibase 10.1063/5.0156008/2902632} {\bibfield  {journal} {\bibinfo  {journal} {J. Chem. Phys.}\ }\textbf {\bibinfo {volume} {159}},\ \bibinfo {pages} {034701} (\bibinfo {year} {2023})}\BibitemShut {NoStop}%
\bibitem [{\citenamefont {P{\'{e}}rez-S{\'{a}}nchez}\ \emph {et~al.}(2023)\citenamefont {P{\'{e}}rez-S{\'{a}}nchez}, \citenamefont {Koner}, \citenamefont {Stern},\ and\ \citenamefont {Yuen-Zhou}}]{Perez-Sanchez2023}%
  \BibitemOpen
  \bibfield  {author} {\bibinfo {author} {\bibfnamefont {J.~B.}\ \bibnamefont {P{\'{e}}rez-S{\'{a}}nchez}}, \bibinfo {author} {\bibfnamefont {A.}~\bibnamefont {Koner}}, \bibinfo {author} {\bibfnamefont {N.~P.}\ \bibnamefont {Stern}}, \ and\ \bibinfo {author} {\bibfnamefont {J.}~\bibnamefont {Yuen-Zhou}},\ }\href {\doibase 10.1073/PNAS.2219223120/SUPPL_FILE/PNAS.2219223120.SAPP.PDF} {\bibfield  {journal} {\bibinfo  {journal} {Proc. Natl. Acad. Sci.}\ }\textbf {\bibinfo {volume} {120}},\ \bibinfo {pages} {e2219223120} (\bibinfo {year} {2023})}\BibitemShut {NoStop}%
\bibitem [{\citenamefont {Sukharev}, \citenamefont {Subotnik},\ and\ \citenamefont {Nitzan}(2023)}]{Sukharev2023}%
  \BibitemOpen
  \bibfield  {author} {\bibinfo {author} {\bibfnamefont {M.}~\bibnamefont {Sukharev}}, \bibinfo {author} {\bibfnamefont {J.}~\bibnamefont {Subotnik}}, \ and\ \bibinfo {author} {\bibfnamefont {A.}~\bibnamefont {Nitzan}},\ }\href {\doibase 10.1063/5.0133972/2868724} {\bibfield  {journal} {\bibinfo  {journal} {J. Chem. Phys.}\ }\textbf {\bibinfo {volume} {158}} (\bibinfo {year} {2023}),\ 10.1063/5.0133972/2868724}\BibitemShut {NoStop}%
\bibitem [{\citenamefont {Yu}\ and\ \citenamefont {Bowman}(2024)}]{Yu2024}%
  \BibitemOpen
  \bibfield  {author} {\bibinfo {author} {\bibfnamefont {Q.}~\bibnamefont {Yu}}\ and\ \bibinfo {author} {\bibfnamefont {J.~M.}\ \bibnamefont {Bowman}},\ }\href {\doibase 10.1021/acs.jctc.4c00129} {\bibfield  {journal} {\bibinfo  {journal} {J. Chem. Theory Comput.}\ }\textbf {\bibinfo {volume} {20}},\ \bibinfo {pages} {4278} (\bibinfo {year} {2024})}\BibitemShut {NoStop}%
\bibitem [{\citenamefont {Liu}, \citenamefont {Chen},\ and\ \citenamefont {Dou}(2024)}]{Liu2024}%
  \BibitemOpen
  \bibfield  {author} {\bibinfo {author} {\bibfnamefont {W.}~\bibnamefont {Liu}}, \bibinfo {author} {\bibfnamefont {J.}~\bibnamefont {Chen}}, \ and\ \bibinfo {author} {\bibfnamefont {W.}~\bibnamefont {Dou}},\ }\href {\doibase 10.1021/acs.jpcc.4c02592} {\bibfield  {journal} {\bibinfo  {journal} {J. Phys. Chem. C}\ }\textbf {\bibinfo {volume} {128}},\ \bibinfo {pages} {12544} (\bibinfo {year} {2024})}\BibitemShut {NoStop}%
\bibitem [{\citenamefont {Sharma}\ and\ \citenamefont {Chen}(2024)}]{Sharma2024}%
  \BibitemOpen
  \bibfield  {author} {\bibinfo {author} {\bibfnamefont {S.~K.}\ \bibnamefont {Sharma}}\ and\ \bibinfo {author} {\bibfnamefont {H.~T.}\ \bibnamefont {Chen}},\ }\href {\doibase 10.1063/5.0225434/3311744} {\bibfield  {journal} {\bibinfo  {journal} {J. Chem. Phys.}\ }\textbf {\bibinfo {volume} {161}},\ \bibinfo {pages} {104102} (\bibinfo {year} {2024})}\BibitemShut {NoStop}%
\bibitem [{\citenamefont {Wickramasinghe}, \citenamefont {Amini},\ and\ \citenamefont {Mandal}(2025)}]{Wickramasinghe2025}%
  \BibitemOpen
  \bibfield  {author} {\bibinfo {author} {\bibfnamefont {S.}~\bibnamefont {Wickramasinghe}}, \bibinfo {author} {\bibfnamefont {A.}~\bibnamefont {Amini}}, \ and\ \bibinfo {author} {\bibfnamefont {A.}~\bibnamefont {Mandal}},\ }\href {https://arxiv.org/pdf/2512.03277 http://arxiv.org/abs/2512.03277} {\bibfield  {journal} {\bibinfo  {journal} {arXiv}\ } (\bibinfo {year} {2025})},\ \Eprint {http://arxiv.org/abs/2512.03277} {arXiv:2512.03277} \BibitemShut {NoStop}%
\bibitem [{\citenamefont {Mondal}\ \emph {et~al.}(2025)\citenamefont {Mondal}, \citenamefont {Park}, \citenamefont {Son}, \citenamefont {Vamivakas}, \citenamefont {Cundiff}, \citenamefont {Krauss},\ and\ \citenamefont {Huo}}]{Mondal2025}%
  \BibitemOpen
  \bibfield  {author} {\bibinfo {author} {\bibfnamefont {E.}~\bibnamefont {Mondal}}, \bibinfo {author} {\bibfnamefont {S.}~\bibnamefont {Park}}, \bibinfo {author} {\bibfnamefont {M.}~\bibnamefont {Son}}, \bibinfo {author} {\bibfnamefont {N.}~\bibnamefont {Vamivakas}}, \bibinfo {author} {\bibfnamefont {S.}~\bibnamefont {Cundiff}}, \bibinfo {author} {\bibfnamefont {T.}~\bibnamefont {Krauss}}, \ and\ \bibinfo {author} {\bibfnamefont {P.}~\bibnamefont {Huo}},\ }\href {\doibase 10.26434/chemrxiv-2025-s1gh5} {\bibfield  {journal} {\bibinfo  {journal} {ChemrXiv}\ } (\bibinfo {year} {2025}),\ 10.26434/chemrxiv-2025-s1gh5}\BibitemShut {NoStop}%
\bibitem [{\citenamefont {Li}, \citenamefont {Subotnik},\ and\ \citenamefont {Nitzan}(2020)}]{liCavityMolecularDynamics2020}%
  \BibitemOpen
  \bibfield  {author} {\bibinfo {author} {\bibfnamefont {T.~E.}\ \bibnamefont {Li}}, \bibinfo {author} {\bibfnamefont {J.~E.}\ \bibnamefont {Subotnik}}, \ and\ \bibinfo {author} {\bibfnamefont {A.}~\bibnamefont {Nitzan}},\ }\href {\doibase 10.1073/pnas.2009272117} {\bibfield  {journal} {\bibinfo  {journal} {Proc. Natl. Acad. Sci. U.S.A.}\ }\textbf {\bibinfo {volume} {117}},\ \bibinfo {pages} {18324} (\bibinfo {year} {2020})}\BibitemShut {NoStop}%
\bibitem [{\citenamefont {Li}, \citenamefont {Nitzan},\ and\ \citenamefont {Subotnik}(2021)}]{liCavityMolecularDynamics2021}%
  \BibitemOpen
  \bibfield  {author} {\bibinfo {author} {\bibfnamefont {T.~E.}\ \bibnamefont {Li}}, \bibinfo {author} {\bibfnamefont {A.}~\bibnamefont {Nitzan}}, \ and\ \bibinfo {author} {\bibfnamefont {J.~E.}\ \bibnamefont {Subotnik}},\ }\href {\doibase 10.1063/5.0037623} {\bibfield  {journal} {\bibinfo  {journal} {J. Chem. Phys.}\ }\textbf {\bibinfo {volume} {154}},\ \bibinfo {pages} {094124} (\bibinfo {year} {2021})}\BibitemShut {NoStop}%
\bibitem [{\citenamefont {Li}, \citenamefont {Nitzan},\ and\ \citenamefont {Subotnik}(2022{\natexlab{a}})}]{liPolaritonRelaxationVibrational2022}%
  \BibitemOpen
  \bibfield  {author} {\bibinfo {author} {\bibfnamefont {T.~E.}\ \bibnamefont {Li}}, \bibinfo {author} {\bibfnamefont {A.}~\bibnamefont {Nitzan}}, \ and\ \bibinfo {author} {\bibfnamefont {J.~E.}\ \bibnamefont {Subotnik}},\ }\href {\doibase 10.1063/5.0079784} {\bibfield  {journal} {\bibinfo  {journal} {J. Chem. Phys.}\ }\textbf {\bibinfo {volume} {156}},\ \bibinfo {pages} {134106} (\bibinfo {year} {2022}{\natexlab{a}})}\BibitemShut {NoStop}%
\bibitem [{\citenamefont {Ji}\ and\ \citenamefont {Li}(2025)}]{jiSelectiveExcitationIRInactive2025}%
  \BibitemOpen
  \bibfield  {author} {\bibinfo {author} {\bibfnamefont {X.}~\bibnamefont {Ji}}\ and\ \bibinfo {author} {\bibfnamefont {T.~E.}\ \bibnamefont {Li}},\ }\href {\doibase 10.1021/acs.jpclett.5c00848} {\bibfield  {journal} {\bibinfo  {journal} {J. Phys. Chem. Lett.}\ }\textbf {\bibinfo {volume} {16}},\ \bibinfo {pages} {5034} (\bibinfo {year} {2025})}\BibitemShut {NoStop}%
\bibitem [{\citenamefont {Li}\ and\ \citenamefont {{Hammes-Schiffer}}(2023)}]{liQMMMModeling2023}%
  \BibitemOpen
  \bibfield  {author} {\bibinfo {author} {\bibfnamefont {T.~E.}\ \bibnamefont {Li}}\ and\ \bibinfo {author} {\bibfnamefont {S.}~\bibnamefont {{Hammes-Schiffer}}},\ }\href {\doibase 10.1021/jacs.2c10170} {\bibfield  {journal} {\bibinfo  {journal} {J. Am. Chem. Soc.}\ }\textbf {\bibinfo {volume} {145}},\ \bibinfo {pages} {377} (\bibinfo {year} {2023})}\BibitemShut {NoStop}%
\bibitem [{\citenamefont {Li}\ \emph {et~al.}(2022)\citenamefont {Li}, \citenamefont {Nitzan}, \citenamefont {{Hammes-Schiffer}},\ and\ \citenamefont {Subotnik}}]{liQuantumSimulationsVibrational2022}%
  \BibitemOpen
  \bibfield  {author} {\bibinfo {author} {\bibfnamefont {T.~E.}\ \bibnamefont {Li}}, \bibinfo {author} {\bibfnamefont {A.}~\bibnamefont {Nitzan}}, \bibinfo {author} {\bibfnamefont {S.}~\bibnamefont {{Hammes-Schiffer}}}, \ and\ \bibinfo {author} {\bibfnamefont {J.~E.}\ \bibnamefont {Subotnik}},\ }\href {\doibase 10.1021/acs.jpclett.2c00613} {\bibfield  {journal} {\bibinfo  {journal} {J. Phys. Chem. Lett.}\ }\textbf {\bibinfo {volume} {13}},\ \bibinfo {pages} {3890} (\bibinfo {year} {2022})}\BibitemShut {NoStop}%
\bibitem [{\citenamefont {Li}, \citenamefont {Nitzan},\ and\ \citenamefont {Subotnik}(2022{\natexlab{b}})}]{liEnergyefficientPathwaySelectively2022}%
  \BibitemOpen
  \bibfield  {author} {\bibinfo {author} {\bibfnamefont {T.~E.}\ \bibnamefont {Li}}, \bibinfo {author} {\bibfnamefont {A.}~\bibnamefont {Nitzan}}, \ and\ \bibinfo {author} {\bibfnamefont {J.~E.}\ \bibnamefont {Subotnik}},\ }\href {\doibase 10.1038/s41467-022-31703-8} {\bibfield  {journal} {\bibinfo  {journal} {Nat. Commun.}\ }\textbf {\bibinfo {volume} {13}},\ \bibinfo {pages} {4203} (\bibinfo {year} {2022}{\natexlab{b}})}\BibitemShut {NoStop}%
\bibitem [{\citenamefont {Li}(2024)}]{liMesoscaleMolecularSimulations2024a}%
  \BibitemOpen
  \bibfield  {author} {\bibinfo {author} {\bibfnamefont {T.~E.}\ \bibnamefont {Li}},\ }\href {\doibase 10.1021/acs.jctc.4c00349} {\bibfield  {journal} {\bibinfo  {journal} {J. Chem. Theory Comput.}\ }\textbf {\bibinfo {volume} {20}},\ \bibinfo {pages} {7016} (\bibinfo {year} {2024})}\BibitemShut {NoStop}%
\bibitem [{\citenamefont {Lieberherr}\ \emph {et~al.}(2023)\citenamefont {Lieberherr}, \citenamefont {Furniss}, \citenamefont {Lawrence},\ and\ \citenamefont {Manolopoulos}}]{lieberherrVibrationalStrongCoupling2023}%
  \BibitemOpen
  \bibfield  {author} {\bibinfo {author} {\bibfnamefont {A.~Z.}\ \bibnamefont {Lieberherr}}, \bibinfo {author} {\bibfnamefont {S.~T.~E.}\ \bibnamefont {Furniss}}, \bibinfo {author} {\bibfnamefont {J.~E.}\ \bibnamefont {Lawrence}}, \ and\ \bibinfo {author} {\bibfnamefont {D.~E.}\ \bibnamefont {Manolopoulos}},\ }\href {\doibase 10.1063/5.0156808} {\bibfield  {journal} {\bibinfo  {journal} {J. Chem. Phys.}\ }\textbf {\bibinfo {volume} {158}},\ \bibinfo {pages} {234106} (\bibinfo {year} {2023})}\BibitemShut {NoStop}%
\bibitem [{\citenamefont {Bowles}\ \emph {et~al.}(2025)\citenamefont {Bowles}, \citenamefont {De~La Fuente~Diez}, \citenamefont {Laage}, \citenamefont {Richardi}, \citenamefont {Vuilleumier},\ and\ \citenamefont {Spezia}}]{bowlesLiquidWaterVibrational2025}%
  \BibitemOpen
  \bibfield  {author} {\bibinfo {author} {\bibfnamefont {J.}~\bibnamefont {Bowles}}, \bibinfo {author} {\bibfnamefont {J.}~\bibnamefont {De~La Fuente~Diez}}, \bibinfo {author} {\bibfnamefont {D.}~\bibnamefont {Laage}}, \bibinfo {author} {\bibfnamefont {J.}~\bibnamefont {Richardi}}, \bibinfo {author} {\bibfnamefont {R.}~\bibnamefont {Vuilleumier}}, \ and\ \bibinfo {author} {\bibfnamefont {R.}~\bibnamefont {Spezia}},\ }\href {\doibase 10.1063/5.0274246} {\bibfield  {journal} {\bibinfo  {journal} {J. Chem. Phys.}\ }\textbf {\bibinfo {volume} {163}},\ \bibinfo {pages} {044104} (\bibinfo {year} {2025})}\BibitemShut {NoStop}%
\bibitem [{\citenamefont {Hasegawa}\ and\ \citenamefont {Tanimura}(2006)}]{hasegawaCalculatingFifthorderRaman2006}%
  \BibitemOpen
  \bibfield  {author} {\bibinfo {author} {\bibfnamefont {T.}~\bibnamefont {Hasegawa}}\ and\ \bibinfo {author} {\bibfnamefont {Y.}~\bibnamefont {Tanimura}},\ }\href {\doibase 10.1063/1.2217947} {\bibfield  {journal} {\bibinfo  {journal} {J. Chem. Phys.}\ }\textbf {\bibinfo {volume} {125}},\ \bibinfo {pages} {074512} (\bibinfo {year} {2006})}\BibitemShut {NoStop}%
\bibitem [{\citenamefont {Ito}, \citenamefont {Hasegawa},\ and\ \citenamefont {Tanimura}(2014)}]{itoCalculatingTwodimensionalTHzRamanTHz2014}%
  \BibitemOpen
  \bibfield  {author} {\bibinfo {author} {\bibfnamefont {H.}~\bibnamefont {Ito}}, \bibinfo {author} {\bibfnamefont {T.}~\bibnamefont {Hasegawa}}, \ and\ \bibinfo {author} {\bibfnamefont {Y.}~\bibnamefont {Tanimura}},\ }\href {\doibase 10.1063/1.4895908} {\bibfield  {journal} {\bibinfo  {journal} {J. Chem. Phys.}\ }\textbf {\bibinfo {volume} {141}},\ \bibinfo {pages} {124503} (\bibinfo {year} {2014})}\BibitemShut {NoStop}%
\bibitem [{\citenamefont {Ito}\ and\ \citenamefont {Tanimura}(2016)}]{itoSimulatingTwoDimensional2016}%
  \BibitemOpen
  \bibfield  {author} {\bibinfo {author} {\bibfnamefont {H.}~\bibnamefont {Ito}}\ and\ \bibinfo {author} {\bibfnamefont {Y.}~\bibnamefont {Tanimura}},\ }\href {\doibase 10.1063/1.4941842} {\bibfield  {journal} {\bibinfo  {journal} {J. Chem. Phys.}\ }\textbf {\bibinfo {volume} {144}},\ \bibinfo {pages} {074201} (\bibinfo {year} {2016})}\BibitemShut {NoStop}%
\bibitem [{\citenamefont {Ciardi}\ \emph {et~al.}(2019)\citenamefont {Ciardi}, \citenamefont {Berger}, \citenamefont {Hamm},\ and\ \citenamefont {Shalit}}]{Ciardi2019}%
  \BibitemOpen
  \bibfield  {author} {\bibinfo {author} {\bibfnamefont {G.}~\bibnamefont {Ciardi}}, \bibinfo {author} {\bibfnamefont {A.}~\bibnamefont {Berger}}, \bibinfo {author} {\bibfnamefont {P.}~\bibnamefont {Hamm}}, \ and\ \bibinfo {author} {\bibfnamefont {A.}~\bibnamefont {Shalit}},\ }\href {\doibase 10.1021/acs.jpclett.9b01528} {\bibfield  {journal} {\bibinfo  {journal} {J. Phys. Chem. Lett.}\ }\textbf {\bibinfo {volume} {10}},\ \bibinfo {pages} {4463} (\bibinfo {year} {2019})}\BibitemShut {NoStop}%
\bibitem [{\citenamefont {Shalit}, \citenamefont {Mousavi},\ and\ \citenamefont {Hamm}(2021)}]{Shalit2DRamanTHz2021}%
  \BibitemOpen
  \bibfield  {author} {\bibinfo {author} {\bibfnamefont {A.}~\bibnamefont {Shalit}}, \bibinfo {author} {\bibfnamefont {S.~J.}\ \bibnamefont {Mousavi}}, \ and\ \bibinfo {author} {\bibfnamefont {P.}~\bibnamefont {Hamm}},\ }\href {\doibase 10.1021/acs.jpcb.0c08962} {\bibfield  {journal} {\bibinfo  {journal} {J. Phys. Chem. B}\ }\textbf {\bibinfo {volume} {125}},\ \bibinfo {pages} {581} (\bibinfo {year} {2021})}\BibitemShut {NoStop}%
\bibitem [{\citenamefont {Mousavi}\ \emph {et~al.}(2022)\citenamefont {Mousavi}, \citenamefont {Berger}, \citenamefont {Hamm},\ and\ \citenamefont {Shalit}}]{MousaviLowFrequencyAnharmonic2022}%
  \BibitemOpen
  \bibfield  {author} {\bibinfo {author} {\bibfnamefont {S.~J.}\ \bibnamefont {Mousavi}}, \bibinfo {author} {\bibfnamefont {A.}~\bibnamefont {Berger}}, \bibinfo {author} {\bibfnamefont {P.}~\bibnamefont {Hamm}}, \ and\ \bibinfo {author} {\bibfnamefont {A.}~\bibnamefont {Shalit}},\ }\href {\doibase 10.1063/5.0090520} {\bibfield  {journal} {\bibinfo  {journal} {J. Chem. Phys.}\ }\textbf {\bibinfo {volume} {156}},\ \bibinfo {pages} {174501} (\bibinfo {year} {2022})}\BibitemShut {NoStop}%
\bibitem [{\citenamefont {Lin}, \citenamefont {Mead},\ and\ \citenamefont {Blake}(2022)}]{LinMapping2022}%
  \BibitemOpen
  \bibfield  {author} {\bibinfo {author} {\bibfnamefont {H.-W.}\ \bibnamefont {Lin}}, \bibinfo {author} {\bibfnamefont {G.}~\bibnamefont {Mead}}, \ and\ \bibinfo {author} {\bibfnamefont {G.~A.}\ \bibnamefont {Blake}},\ }\href {\doibase 10.1103/PhysRevLett.129.207401} {\bibfield  {journal} {\bibinfo  {journal} {Phys. Rev. Lett.}\ }\textbf {\bibinfo {volume} {129}},\ \bibinfo {pages} {207401} (\bibinfo {year} {2022})}\BibitemShut {NoStop}%
\bibitem [{\citenamefont {Seliya}, \citenamefont {Bonn},\ and\ \citenamefont {Grechko}(2024)}]{SeliyaOnSelectionRules2024}%
  \BibitemOpen
  \bibfield  {author} {\bibinfo {author} {\bibfnamefont {P.}~\bibnamefont {Seliya}}, \bibinfo {author} {\bibfnamefont {M.}~\bibnamefont {Bonn}}, \ and\ \bibinfo {author} {\bibfnamefont {M.}~\bibnamefont {Grechko}},\ }\href {\doibase 10.1063/5.0179041} {\bibfield  {journal} {\bibinfo  {journal} {J. Chem. Phys.}\ }\textbf {\bibinfo {volume} {160}},\ \bibinfo {pages} {034201} (\bibinfo {year} {2024})}\BibitemShut {NoStop}%
\bibitem [{\citenamefont {Begu{\v s}i{\'c}}\ and\ \citenamefont {Blake}(2023)}]{begusicTwodimensionalInfraredRamanSpectroscopy2023}%
  \BibitemOpen
  \bibfield  {author} {\bibinfo {author} {\bibfnamefont {T.}~\bibnamefont {Begu{\v s}i{\'c}}}\ and\ \bibinfo {author} {\bibfnamefont {G.~A.}\ \bibnamefont {Blake}},\ }\href {\doibase 10.1038/s41467-023-37667-7} {\bibfield  {journal} {\bibinfo  {journal} {Nat. Commun.}\ }\textbf {\bibinfo {volume} {14}},\ \bibinfo {pages} {1950} (\bibinfo {year} {2023})}\BibitemShut {NoStop}%
\bibitem [{\citenamefont {Begu{\v s}i{\'c}}\ \emph {et~al.}(2022)\citenamefont {Begu{\v s}i{\'c}}, \citenamefont {Tao}, \citenamefont {Blake},\ and\ \citenamefont {Miller}}]{begusicEquilibriumNonequilibriumRingpolymer2022}%
  \BibitemOpen
  \bibfield  {author} {\bibinfo {author} {\bibfnamefont {T.}~\bibnamefont {Begu{\v s}i{\'c}}}, \bibinfo {author} {\bibfnamefont {X.}~\bibnamefont {Tao}}, \bibinfo {author} {\bibfnamefont {G.~A.}\ \bibnamefont {Blake}}, \ and\ \bibinfo {author} {\bibfnamefont {T.~F.}\ \bibnamefont {Miller}},\ }\href {\doibase 10.1063/5.0087156} {\bibfield  {journal} {\bibinfo  {journal} {J. Chem. Phys.}\ }\textbf {\bibinfo {volume} {156}},\ \bibinfo {pages} {131102} (\bibinfo {year} {2022})}\BibitemShut {NoStop}%
\bibitem [{\citenamefont {Medders}\ and\ \citenamefont {Paesani}(2015{\natexlab{a}})}]{Medders2015}%
  \BibitemOpen
  \bibfield  {author} {\bibinfo {author} {\bibfnamefont {G.~R.}\ \bibnamefont {Medders}}\ and\ \bibinfo {author} {\bibfnamefont {F.}~\bibnamefont {Paesani}},\ }\href {\doibase 10.1063/1.4916629/962952} {\bibfield  {journal} {\bibinfo  {journal} {J. Chem. Phys.}\ }\textbf {\bibinfo {volume} {142}},\ \bibinfo {pages} {212411} (\bibinfo {year} {2015}{\natexlab{a}})}\BibitemShut {NoStop}%
\bibitem [{\citenamefont {Sommers}\ \emph {et~al.}(2020)\citenamefont {Sommers}, \citenamefont {{Calegari Andrade}}, \citenamefont {Zhang}, \citenamefont {Wang},\ and\ \citenamefont {Car}}]{Sommers2020}%
  \BibitemOpen
  \bibfield  {author} {\bibinfo {author} {\bibfnamefont {G.~M.}\ \bibnamefont {Sommers}}, \bibinfo {author} {\bibfnamefont {M.~F.}\ \bibnamefont {{Calegari Andrade}}}, \bibinfo {author} {\bibfnamefont {L.}~\bibnamefont {Zhang}}, \bibinfo {author} {\bibfnamefont {H.}~\bibnamefont {Wang}}, \ and\ \bibinfo {author} {\bibfnamefont {R.}~\bibnamefont {Car}},\ }\href {\doibase 10.1039/D0CP01893G} {\bibfield  {journal} {\bibinfo  {journal} {Phys. Chem. Chem. Phys.}\ }\textbf {\bibinfo {volume} {22}},\ \bibinfo {pages} {10592} (\bibinfo {year} {2020})}\BibitemShut {NoStop}%
\bibitem [{\citenamefont {Omodemi}\ and\ \citenamefont {Kaledin}(2025)}]{Omodemi2025}%
  \BibitemOpen
  \bibfield  {author} {\bibinfo {author} {\bibfnamefont {O.}~\bibnamefont {Omodemi}}\ and\ \bibinfo {author} {\bibfnamefont {M.}~\bibnamefont {Kaledin}},\ }\href {\doibase 10.1021/acs.jpca.5c05184} {\bibfield  {journal} {\bibinfo  {journal} {J. Phys. Chem. A}\ }\textbf {\bibinfo {volume} {129}},\ \bibinfo {pages} {9921} (\bibinfo {year} {2025})}\BibitemShut {NoStop}%
\bibitem [{\citenamefont {Carusotto}\ and\ \citenamefont {Ciuti}(2013)}]{Carusotto2013}%
  \BibitemOpen
  \bibfield  {author} {\bibinfo {author} {\bibfnamefont {I.}~\bibnamefont {Carusotto}}\ and\ \bibinfo {author} {\bibfnamefont {C.}~\bibnamefont {Ciuti}},\ }\href {\doibase 10.1103/RevModPhys.85.299} {\bibfield  {journal} {\bibinfo  {journal} {Rev. Mod. Phys.}\ }\textbf {\bibinfo {volume} {85}},\ \bibinfo {pages} {299} (\bibinfo {year} {2013})}\BibitemShut {NoStop}%
\bibitem [{\citenamefont {Szidarovszky}(2023)}]{szidarovszkyEfficientFlexibleApproach2023}%
  \BibitemOpen
  \bibfield  {author} {\bibinfo {author} {\bibfnamefont {T.}~\bibnamefont {Szidarovszky}},\ }\href {\doibase 10.1063/5.0153293} {\bibfield  {journal} {\bibinfo  {journal} {J. Chem. Phys.}\ }\textbf {\bibinfo {volume} {159}},\ \bibinfo {pages} {014112} (\bibinfo {year} {2023})}\BibitemShut {NoStop}%
\bibitem [{\citenamefont {Schnappinger}\ and\ \citenamefont {Kowalewski}(2025)}]{schnappingerMolecularPolarizabilityVibrational2025}%
  \BibitemOpen
  \bibfield  {author} {\bibinfo {author} {\bibfnamefont {T.}~\bibnamefont {Schnappinger}}\ and\ \bibinfo {author} {\bibfnamefont {M.}~\bibnamefont {Kowalewski}},\ }\href {\doibase 10.1021/acs.jctc.5c00461} {\bibfield  {journal} {\bibinfo  {journal} {J. Chem. Theory Comput.}\ }\textbf {\bibinfo {volume} {21}},\ \bibinfo {pages} {5171} (\bibinfo {year} {2025})}\BibitemShut {NoStop}%
\bibitem [{\citenamefont {Li}, \citenamefont {Nitzan},\ and\ \citenamefont {Subotnik}(2020)}]{Li2020Origin}%
  \BibitemOpen
  \bibfield  {author} {\bibinfo {author} {\bibfnamefont {T.~E.}\ \bibnamefont {Li}}, \bibinfo {author} {\bibfnamefont {A.}~\bibnamefont {Nitzan}}, \ and\ \bibinfo {author} {\bibfnamefont {J.~E.}\ \bibnamefont {Subotnik}},\ }\href {\doibase 10.1063/5.0006472} {\bibfield  {journal} {\bibinfo  {journal} {J. Chem. Phys.}\ }\textbf {\bibinfo {volume} {152}},\ \bibinfo {pages} {234107} (\bibinfo {year} {2020})}\BibitemShut {NoStop}%
\bibitem [{\citenamefont {Habershon}, \citenamefont {Markland},\ and\ \citenamefont {Manolopoulos}(2009)}]{habershonCompetingQuantumEffects2009}%
  \BibitemOpen
  \bibfield  {author} {\bibinfo {author} {\bibfnamefont {S.}~\bibnamefont {Habershon}}, \bibinfo {author} {\bibfnamefont {T.~E.}\ \bibnamefont {Markland}}, \ and\ \bibinfo {author} {\bibfnamefont {D.~E.}\ \bibnamefont {Manolopoulos}},\ }\href {\doibase 10.1063/1.3167790} {\bibfield  {journal} {\bibinfo  {journal} {J. Chem. Phys.}\ }\textbf {\bibinfo {volume} {131}},\ \bibinfo {pages} {024501} (\bibinfo {year} {2009})}\BibitemShut {NoStop}%
\bibitem [{\citenamefont {Hamm}(2014)}]{hamm2DRamanTHzSpectroscopySensitive2014}%
  \BibitemOpen
  \bibfield  {author} {\bibinfo {author} {\bibfnamefont {P.}~\bibnamefont {Hamm}},\ }\href {\doibase 10.1063/1.4901216} {\bibfield  {journal} {\bibinfo  {journal} {J. Chem. Phys.}\ }\textbf {\bibinfo {volume} {141}},\ \bibinfo {pages} {184201} (\bibinfo {year} {2014})}\BibitemShut {NoStop}%
\bibitem [{\citenamefont {Heyden}\ \emph {et~al.}(2010)\citenamefont {Heyden}, \citenamefont {Sun}, \citenamefont {Funkner}, \citenamefont {Mathias}, \citenamefont {Forbert}, \citenamefont {Havenith},\ and\ \citenamefont {Marx}}]{HeydenAbInitioMolecularDynamics2010}%
  \BibitemOpen
  \bibfield  {author} {\bibinfo {author} {\bibfnamefont {M.}~\bibnamefont {Heyden}}, \bibinfo {author} {\bibfnamefont {J.}~\bibnamefont {Sun}}, \bibinfo {author} {\bibfnamefont {S.}~\bibnamefont {Funkner}}, \bibinfo {author} {\bibfnamefont {G.}~\bibnamefont {Mathias}}, \bibinfo {author} {\bibfnamefont {H.}~\bibnamefont {Forbert}}, \bibinfo {author} {\bibfnamefont {M.}~\bibnamefont {Havenith}}, \ and\ \bibinfo {author} {\bibfnamefont {D.}~\bibnamefont {Marx}},\ }\href {\doibase 10.1073/pnas.0914885107} {\bibfield  {journal} {\bibinfo  {journal} {Proc. Natl. Acad. Sci. U. S. A.}\ }\textbf {\bibinfo {volume} {107}},\ \bibinfo {pages} {12068} (\bibinfo {year} {2010})}\BibitemShut {NoStop}%
\bibitem [{\citenamefont {Wan}\ \emph {et~al.}(2013)\citenamefont {Wan}, \citenamefont {Spanu}, \citenamefont {Galli},\ and\ \citenamefont {Gygi}}]{WanRamanSpectraOfLiquidWater2013}%
  \BibitemOpen
  \bibfield  {author} {\bibinfo {author} {\bibfnamefont {Q.}~\bibnamefont {Wan}}, \bibinfo {author} {\bibfnamefont {L.}~\bibnamefont {Spanu}}, \bibinfo {author} {\bibfnamefont {G.~A.}\ \bibnamefont {Galli}}, \ and\ \bibinfo {author} {\bibfnamefont {F.}~\bibnamefont {Gygi}},\ }\href {\doibase 10.1021/ct4005307} {\bibfield  {journal} {\bibinfo  {journal} {J. Chem. Theory Comput.}\ }\textbf {\bibinfo {volume} {9}},\ \bibinfo {pages} {4124} (\bibinfo {year} {2013})}\BibitemShut {NoStop}%
\bibitem [{\citenamefont {Marsalek}\ and\ \citenamefont {Markland}(2017)}]{MarsalekQuantumDynamicsAndSpectroscopy2017}%
  \BibitemOpen
  \bibfield  {author} {\bibinfo {author} {\bibfnamefont {O.}~\bibnamefont {Marsalek}}\ and\ \bibinfo {author} {\bibfnamefont {T.~E.}\ \bibnamefont {Markland}},\ }\href {\doibase 10.1021/acs.jpclett.7b00391} {\bibfield  {journal} {\bibinfo  {journal} {J. Phys. Chem. Lett.}\ }\textbf {\bibinfo {volume} {8}},\ \bibinfo {pages} {1545} (\bibinfo {year} {2017})}\BibitemShut {NoStop}%
\bibitem [{\citenamefont {Rozsa}\ \emph {et~al.}(2018)\citenamefont {Rozsa}, \citenamefont {Pan}, \citenamefont {Giberti},\ and\ \citenamefont {Galli}}]{RozsaAbInitioSpectroscopy2018}%
  \BibitemOpen
  \bibfield  {author} {\bibinfo {author} {\bibfnamefont {V.}~\bibnamefont {Rozsa}}, \bibinfo {author} {\bibfnamefont {D.}~\bibnamefont {Pan}}, \bibinfo {author} {\bibfnamefont {F.}~\bibnamefont {Giberti}}, \ and\ \bibinfo {author} {\bibfnamefont {G.}~\bibnamefont {Galli}},\ }\href {\doibase 10.1073/pnas.1800123115} {\bibfield  {journal} {\bibinfo  {journal} {Proc. Natl. Acad. Sci. U. S. A.}\ }\textbf {\bibinfo {volume} {115}},\ \bibinfo {pages} {6952} (\bibinfo {year} {2018})}\BibitemShut {NoStop}%
\bibitem [{\citenamefont {Ruiz~Pestana}\ \emph {et~al.}(2018)\citenamefont {Ruiz~Pestana}, \citenamefont {Marsalek}, \citenamefont {Markland},\ and\ \citenamefont {Head-Gordon}}]{PestanaTheQuestForAccurate2018}%
  \BibitemOpen
  \bibfield  {author} {\bibinfo {author} {\bibfnamefont {L.}~\bibnamefont {Ruiz~Pestana}}, \bibinfo {author} {\bibfnamefont {O.}~\bibnamefont {Marsalek}}, \bibinfo {author} {\bibfnamefont {T.~E.}\ \bibnamefont {Markland}}, \ and\ \bibinfo {author} {\bibfnamefont {T.}~\bibnamefont {Head-Gordon}},\ }\href {\doibase 10.1021/acs.jpclett.8b02400} {\bibfield  {journal} {\bibinfo  {journal} {J. Phys. Chem. Lett.}\ }\textbf {\bibinfo {volume} {9}},\ \bibinfo {pages} {5009} (\bibinfo {year} {2018})}\BibitemShut {NoStop}%
\bibitem [{\citenamefont {Cassone}\ \emph {et~al.}(2019)\citenamefont {Cassone}, \citenamefont {Sponer}, \citenamefont {Trusso},\ and\ \citenamefont {Saija}}]{CassoneAbInitioSpectroscopy2019}%
  \BibitemOpen
  \bibfield  {author} {\bibinfo {author} {\bibfnamefont {G.}~\bibnamefont {Cassone}}, \bibinfo {author} {\bibfnamefont {J.}~\bibnamefont {Sponer}}, \bibinfo {author} {\bibfnamefont {S.}~\bibnamefont {Trusso}}, \ and\ \bibinfo {author} {\bibfnamefont {F.}~\bibnamefont {Saija}},\ }\href {\doibase 10.1039/C9CP03101D} {\bibfield  {journal} {\bibinfo  {journal} {Phys. Chem. Chem. Phys.}\ }\textbf {\bibinfo {volume} {21}},\ \bibinfo {pages} {21205} (\bibinfo {year} {2019})}\BibitemShut {NoStop}%
\bibitem [{\citenamefont {Ohto}\ \emph {et~al.}(2019)\citenamefont {Ohto}, \citenamefont {Dodia}, \citenamefont {Xu}, \citenamefont {Imoto}, \citenamefont {Tang}, \citenamefont {Zysk}, \citenamefont {K{\"u}hne}, \citenamefont {Shigeta}, \citenamefont {Bonn}, \citenamefont {Wu},\ and\ \citenamefont {Nagata}}]{OhtoAccessingTheAccuracy2019}%
  \BibitemOpen
  \bibfield  {author} {\bibinfo {author} {\bibfnamefont {T.}~\bibnamefont {Ohto}}, \bibinfo {author} {\bibfnamefont {M.}~\bibnamefont {Dodia}}, \bibinfo {author} {\bibfnamefont {J.}~\bibnamefont {Xu}}, \bibinfo {author} {\bibfnamefont {S.}~\bibnamefont {Imoto}}, \bibinfo {author} {\bibfnamefont {F.}~\bibnamefont {Tang}}, \bibinfo {author} {\bibfnamefont {F.}~\bibnamefont {Zysk}}, \bibinfo {author} {\bibfnamefont {T.~D.}\ \bibnamefont {K{\"u}hne}}, \bibinfo {author} {\bibfnamefont {Y.}~\bibnamefont {Shigeta}}, \bibinfo {author} {\bibfnamefont {M.}~\bibnamefont {Bonn}}, \bibinfo {author} {\bibfnamefont {X.}~\bibnamefont {Wu}}, \ and\ \bibinfo {author} {\bibfnamefont {Y.}~\bibnamefont {Nagata}},\ }\href {\doibase 10.1021/acs.jpclett.9b01983} {\bibfield  {journal} {\bibinfo  {journal} {J. Phys. Chem. Lett.}\ }\textbf {\bibinfo {volume} {10}},\ \bibinfo {pages} {4914} (\bibinfo {year} {2019})}\BibitemShut {NoStop}%
\bibitem [{\citenamefont {Hasegawa}\ and\ \citenamefont {Tanimura}(2011)}]{HasegawaAPolarizableWaterModel2011}%
  \BibitemOpen
  \bibfield  {author} {\bibinfo {author} {\bibfnamefont {T.}~\bibnamefont {Hasegawa}}\ and\ \bibinfo {author} {\bibfnamefont {Y.}~\bibnamefont {Tanimura}},\ }\href {\doibase 10.1021/jp111308f} {\bibfield  {journal} {\bibinfo  {journal} {J. Phys. Chem. B}\ }\textbf {\bibinfo {volume} {115}},\ \bibinfo {pages} {5545} (\bibinfo {year} {2011})}\BibitemShut {NoStop}%
\bibitem [{\citenamefont {Medders}\ and\ \citenamefont {Paesani}(2015{\natexlab{b}})}]{meddersInfraredRamanSpectroscopy2015}%
  \BibitemOpen
  \bibfield  {author} {\bibinfo {author} {\bibfnamefont {G.~R.}\ \bibnamefont {Medders}}\ and\ \bibinfo {author} {\bibfnamefont {F.}~\bibnamefont {Paesani}},\ }\href {\doibase 10.1021/ct501131j} {\bibfield  {journal} {\bibinfo  {journal} {J. Chem. Theory Comput.}\ }\textbf {\bibinfo {volume} {11}},\ \bibinfo {pages} {1145} (\bibinfo {year} {2015}{\natexlab{b}})}\BibitemShut {NoStop}%
\bibitem [{\citenamefont {Ito}, \citenamefont {Hasegawa},\ and\ \citenamefont {Tanimura}(2016)}]{itoEffectsIntermolecularCharge2016}%
  \BibitemOpen
  \bibfield  {author} {\bibinfo {author} {\bibfnamefont {H.}~\bibnamefont {Ito}}, \bibinfo {author} {\bibfnamefont {T.}~\bibnamefont {Hasegawa}}, \ and\ \bibinfo {author} {\bibfnamefont {Y.}~\bibnamefont {Tanimura}},\ }\href {\doibase 10.1021/acs.jpclett.6b01766} {\bibfield  {journal} {\bibinfo  {journal} {J. Phys. Chem. Lett.}\ }\textbf {\bibinfo {volume} {7}},\ \bibinfo {pages} {4147} (\bibinfo {year} {2016})}\BibitemShut {NoStop}%
\bibitem [{\citenamefont {Sidler}, \citenamefont {Meuwly},\ and\ \citenamefont {Hamm}(2018)}]{sidlerEfficientWaterForce2018}%
  \BibitemOpen
  \bibfield  {author} {\bibinfo {author} {\bibfnamefont {D.}~\bibnamefont {Sidler}}, \bibinfo {author} {\bibfnamefont {M.}~\bibnamefont {Meuwly}}, \ and\ \bibinfo {author} {\bibfnamefont {P.}~\bibnamefont {Hamm}},\ }\href {\doibase 10.1063/1.5037062} {\bibfield  {journal} {\bibinfo  {journal} {J. Chem. Phys.}\ }\textbf {\bibinfo {volume} {148}},\ \bibinfo {pages} {244504} (\bibinfo {year} {2018})}\BibitemShut {NoStop}%
\bibitem [{\citenamefont {Gastegger}, \citenamefont {Behler},\ and\ \citenamefont {Marquetand}(2017)}]{GasteggerMachineLearningMolecularDynamics2017}%
  \BibitemOpen
  \bibfield  {author} {\bibinfo {author} {\bibfnamefont {M.}~\bibnamefont {Gastegger}}, \bibinfo {author} {\bibfnamefont {J.}~\bibnamefont {Behler}}, \ and\ \bibinfo {author} {\bibfnamefont {P.}~\bibnamefont {Marquetand}},\ }\href {\doibase 10.1039/C7SC02267K} {\bibfield  {journal} {\bibinfo  {journal} {Chem. Sci.}\ }\textbf {\bibinfo {volume} {8}},\ \bibinfo {pages} {6924} (\bibinfo {year} {2017})}\BibitemShut {NoStop}%
\bibitem [{\citenamefont {Morawietz}\ \emph {et~al.}(2018)\citenamefont {Morawietz}, \citenamefont {Marsalek}, \citenamefont {Pattenaude}, \citenamefont {Streacker}, \citenamefont {Ben-Amotz},\ and\ \citenamefont {Markland}}]{MorawietzTheInterplayOfStructure2018}%
  \BibitemOpen
  \bibfield  {author} {\bibinfo {author} {\bibfnamefont {T.}~\bibnamefont {Morawietz}}, \bibinfo {author} {\bibfnamefont {O.}~\bibnamefont {Marsalek}}, \bibinfo {author} {\bibfnamefont {S.~R.}\ \bibnamefont {Pattenaude}}, \bibinfo {author} {\bibfnamefont {L.~M.}\ \bibnamefont {Streacker}}, \bibinfo {author} {\bibfnamefont {D.}~\bibnamefont {Ben-Amotz}}, \ and\ \bibinfo {author} {\bibfnamefont {T.~E.}\ \bibnamefont {Markland}},\ }\href {\doibase 10.1021/acs.jpclett.8b00133} {\bibfield  {journal} {\bibinfo  {journal} {J. Phys. Chem. Lett.}\ }\textbf {\bibinfo {volume} {9}},\ \bibinfo {pages} {851} (\bibinfo {year} {2018})}\BibitemShut {NoStop}%
\bibitem [{\citenamefont {Kapil}\ \emph {et~al.}(2020)\citenamefont {Kapil}, \citenamefont {Wilkins}, \citenamefont {Lan},\ and\ \citenamefont {Ceriotti}}]{KapilInexpensiveModelingOfQuantumDynamics2020}%
  \BibitemOpen
  \bibfield  {author} {\bibinfo {author} {\bibfnamefont {V.}~\bibnamefont {Kapil}}, \bibinfo {author} {\bibfnamefont {D.~M.}\ \bibnamefont {Wilkins}}, \bibinfo {author} {\bibfnamefont {J.}~\bibnamefont {Lan}}, \ and\ \bibinfo {author} {\bibfnamefont {M.}~\bibnamefont {Ceriotti}},\ }\href {\doibase 10.1063/1.5141950} {\bibfield  {journal} {\bibinfo  {journal} {J. Chem. Phys.}\ }\textbf {\bibinfo {volume} {152}},\ \bibinfo {pages} {124104} (\bibinfo {year} {2020})}\BibitemShut {NoStop}%
\bibitem [{\citenamefont {Schienbein}(2023)}]{SchienbeinSpectroscopyFromMachineLearning2023}%
  \BibitemOpen
  \bibfield  {author} {\bibinfo {author} {\bibfnamefont {P.}~\bibnamefont {Schienbein}},\ }\href {\doibase 10.1021/acs.jctc.2c00788} {\bibfield  {journal} {\bibinfo  {journal} {J. Chem. Theory Comput.}\ }\textbf {\bibinfo {volume} {19}},\ \bibinfo {pages} {705} (\bibinfo {year} {2023})}\BibitemShut {NoStop}%
\bibitem [{\citenamefont {Inoue}\ \emph {et~al.}(2023)\citenamefont {Inoue}, \citenamefont {Litman}, \citenamefont {Wilkins}, \citenamefont {Nagata},\ and\ \citenamefont {Okuno}}]{InoueIsUnifiedUnderstanding2023}%
  \BibitemOpen
  \bibfield  {author} {\bibinfo {author} {\bibfnamefont {K.}~\bibnamefont {Inoue}}, \bibinfo {author} {\bibfnamefont {Y.}~\bibnamefont {Litman}}, \bibinfo {author} {\bibfnamefont {D.~M.}\ \bibnamefont {Wilkins}}, \bibinfo {author} {\bibfnamefont {Y.}~\bibnamefont {Nagata}}, \ and\ \bibinfo {author} {\bibfnamefont {M.}~\bibnamefont {Okuno}},\ }\href {\doibase 10.1021/acs.jpclett.3c00398} {\bibfield  {journal} {\bibinfo  {journal} {J. Phys. Chem. Lett.}\ }\textbf {\bibinfo {volume} {14}},\ \bibinfo {pages} {3063} (\bibinfo {year} {2023})}\BibitemShut {NoStop}%
\bibitem [{\citenamefont {Kapil}\ \emph {et~al.}(2024)\citenamefont {Kapil}, \citenamefont {Kovács}, \citenamefont {Csányi},\ and\ \citenamefont {Michaelides}}]{KapilFirstPrinciplesSpectroscopy2024}%
  \BibitemOpen
  \bibfield  {author} {\bibinfo {author} {\bibfnamefont {V.}~\bibnamefont {Kapil}}, \bibinfo {author} {\bibfnamefont {D.~P.}\ \bibnamefont {Kovács}}, \bibinfo {author} {\bibfnamefont {G.}~\bibnamefont {Csányi}}, \ and\ \bibinfo {author} {\bibfnamefont {A.}~\bibnamefont {Michaelides}},\ }\href {\doibase 10.1039/D3FD00113J} {\bibfield  {journal} {\bibinfo  {journal} {Faraday Discuss.}\ }\textbf {\bibinfo {volume} {249}},\ \bibinfo {pages} {50} (\bibinfo {year} {2024})}\BibitemShut {NoStop}%
\bibitem [{\citenamefont {Avila}(2005)}]{avilaInitioDipolePolarizability2005}%
  \BibitemOpen
  \bibfield  {author} {\bibinfo {author} {\bibfnamefont {G.}~\bibnamefont {Avila}},\ }\href {\doibase 10.1063/1.1867437} {\bibfield  {journal} {\bibinfo  {journal} {J. Chem. Phys.}\ }\textbf {\bibinfo {volume} {122}},\ \bibinfo {pages} {144310} (\bibinfo {year} {2005})}\BibitemShut {NoStop}%
\bibitem [{\citenamefont {Ito}, \citenamefont {Jo},\ and\ \citenamefont {Tanimura}(2015)}]{itoNotesSimulatingTwodimensional2015}%
  \BibitemOpen
  \bibfield  {author} {\bibinfo {author} {\bibfnamefont {H.}~\bibnamefont {Ito}}, \bibinfo {author} {\bibfnamefont {J.-Y.}\ \bibnamefont {Jo}}, \ and\ \bibinfo {author} {\bibfnamefont {Y.}~\bibnamefont {Tanimura}},\ }\href {\doibase 10.1063/1.4932597} {\bibfield  {journal} {\bibinfo  {journal} {Struct. Dyn.}\ }\textbf {\bibinfo {volume} {2}},\ \bibinfo {pages} {054102} (\bibinfo {year} {2015})}\BibitemShut {NoStop}%
\bibitem [{\citenamefont {Habershon}, \citenamefont {Fanourgakis},\ and\ \citenamefont {Manolopoulos}(2008)}]{habershonComparisonPathIntegral2008}%
  \BibitemOpen
  \bibfield  {author} {\bibinfo {author} {\bibfnamefont {S.}~\bibnamefont {Habershon}}, \bibinfo {author} {\bibfnamefont {G.~S.}\ \bibnamefont {Fanourgakis}}, \ and\ \bibinfo {author} {\bibfnamefont {D.~E.}\ \bibnamefont {Manolopoulos}},\ }\href {\doibase 10.1063/1.2968555} {\bibfield  {journal} {\bibinfo  {journal} {J. Chem. Phys.}\ }\textbf {\bibinfo {volume} {129}},\ \bibinfo {pages} {074501} (\bibinfo {year} {2008})}\BibitemShut {NoStop}%
\bibitem [{\citenamefont {Sun}(2019)}]{sunHybridEquilibriumnonequilibriumMolecular2019}%
  \BibitemOpen
  \bibfield  {author} {\bibinfo {author} {\bibfnamefont {X.}~\bibnamefont {Sun}},\ }\href {\doibase 10.1063/1.5130926} {\bibfield  {journal} {\bibinfo  {journal} {J. Chem. Phys.}\ }\textbf {\bibinfo {volume} {151}},\ \bibinfo {pages} {194507} (\bibinfo {year} {2019})}\BibitemShut {NoStop}%
\bibitem [{\citenamefont {Hamm}\ and\ \citenamefont {Savolainen}(2012)}]{hammTwodimensionalRamanterahertzSpectroscopyWater2012}%
  \BibitemOpen
  \bibfield  {author} {\bibinfo {author} {\bibfnamefont {P.}~\bibnamefont {Hamm}}\ and\ \bibinfo {author} {\bibfnamefont {J.}~\bibnamefont {Savolainen}},\ }\href {\doibase 10.1063/1.3691601} {\bibfield  {journal} {\bibinfo  {journal} {J. Chem. Phys.}\ }\textbf {\bibinfo {volume} {136}},\ \bibinfo {pages} {094516} (\bibinfo {year} {2012})}\BibitemShut {NoStop}%
\bibitem [{\citenamefont {Mart{\'i}nez}\ \emph {et~al.}(2009)\citenamefont {Mart{\'i}nez}, \citenamefont {Andrade}, \citenamefont {Birgin},\ and\ \citenamefont {Mart{\'i}nez}}]{martinezPACKMOLPackageBuilding2009}%
  \BibitemOpen
  \bibfield  {author} {\bibinfo {author} {\bibfnamefont {L.}~\bibnamefont {Mart{\'i}nez}}, \bibinfo {author} {\bibfnamefont {R.}~\bibnamefont {Andrade}}, \bibinfo {author} {\bibfnamefont {E.~G.}\ \bibnamefont {Birgin}}, \ and\ \bibinfo {author} {\bibfnamefont {J.~M.}\ \bibnamefont {Mart{\'i}nez}},\ }\href {\doibase 10.1002/jcc.21224} {\bibfield  {journal} {\bibinfo  {journal} {J. Comput. Chem.}\ }\textbf {\bibinfo {volume} {30}},\ \bibinfo {pages} {2157} (\bibinfo {year} {2009})}\BibitemShut {NoStop}%
\bibitem [{Note1()}]{Note1}%
  \BibitemOpen
  \bibinfo {note} {If the dipole derivatives corresponding to the fixed point-charge dipole model [Eq. \protect \eqref {eq:dipole_derivatives_fixed_charge}] were used to couple to the delta electric pulse, the resulting 2D-IIR spectra were not meaningfully modified.}\BibitemShut {Stop}%
\bibitem [{\citenamefont {Litman}\ \emph {et~al.}(2024)\citenamefont {Litman}, \citenamefont {Kapil}, \citenamefont {Feldman}, \citenamefont {Tisi}, \citenamefont {Begu{\v s}i{\'c}}, \citenamefont {Fidanyan}, \citenamefont {Fraux}, \citenamefont {Higer}, \citenamefont {Kellner}, \citenamefont {Li}, \citenamefont {P{\'o}s}, \citenamefont {Stocco}, \citenamefont {Trenins}, \citenamefont {Hirshberg}, \citenamefont {Rossi},\ and\ \citenamefont {Ceriotti}}]{litmanIPI30Flexible2024}%
  \BibitemOpen
  \bibfield  {author} {\bibinfo {author} {\bibfnamefont {Y.}~\bibnamefont {Litman}}, \bibinfo {author} {\bibfnamefont {V.}~\bibnamefont {Kapil}}, \bibinfo {author} {\bibfnamefont {Y.~M.~Y.}\ \bibnamefont {Feldman}}, \bibinfo {author} {\bibfnamefont {D.}~\bibnamefont {Tisi}}, \bibinfo {author} {\bibfnamefont {T.}~\bibnamefont {Begu{\v s}i{\'c}}}, \bibinfo {author} {\bibfnamefont {K.}~\bibnamefont {Fidanyan}}, \bibinfo {author} {\bibfnamefont {G.}~\bibnamefont {Fraux}}, \bibinfo {author} {\bibfnamefont {J.}~\bibnamefont {Higer}}, \bibinfo {author} {\bibfnamefont {M.}~\bibnamefont {Kellner}}, \bibinfo {author} {\bibfnamefont {T.~E.}\ \bibnamefont {Li}}, \bibinfo {author} {\bibfnamefont {E.~S.}\ \bibnamefont {P{\'o}s}}, \bibinfo {author} {\bibfnamefont {E.}~\bibnamefont {Stocco}}, \bibinfo {author} {\bibfnamefont {G.}~\bibnamefont {Trenins}}, \bibinfo {author} {\bibfnamefont {B.}~\bibnamefont {Hirshberg}}, \bibinfo {author} {\bibfnamefont {M.}~\bibnamefont {Rossi}}, \ and\ \bibinfo {author} {\bibfnamefont
  {M.}~\bibnamefont {Ceriotti}},\ }\href {\doibase 10.1063/5.0215869} {\bibfield  {journal} {\bibinfo  {journal} {J. Chem. Phys.}\ }\textbf {\bibinfo {volume} {161}},\ \bibinfo {pages} {062504} (\bibinfo {year} {2024})}\BibitemShut {NoStop}%
\bibitem [{\citenamefont {Houdr{\'{e}}}, \citenamefont {Stanley},\ and\ \citenamefont {Ilegems}(1996)}]{Houdre1996}%
  \BibitemOpen
  \bibfield  {author} {\bibinfo {author} {\bibfnamefont {R.}~\bibnamefont {Houdr{\'{e}}}}, \bibinfo {author} {\bibfnamefont {R.~P.}\ \bibnamefont {Stanley}}, \ and\ \bibinfo {author} {\bibfnamefont {M.}~\bibnamefont {Ilegems}},\ }\href {\doibase 10.1103/PhysRevA.53.2711} {\bibfield  {journal} {\bibinfo  {journal} {Phys. Rev. A}\ }\textbf {\bibinfo {volume} {53}},\ \bibinfo {pages} {2711} (\bibinfo {year} {1996})}\BibitemShut {NoStop}%
\bibitem [{\citenamefont {del Pino}, \citenamefont {Feist},\ and\ \citenamefont {Garcia-Vidal}(2015)}]{DelPino2015}%
  \BibitemOpen
  \bibfield  {author} {\bibinfo {author} {\bibfnamefont {J.}~\bibnamefont {del Pino}}, \bibinfo {author} {\bibfnamefont {J.}~\bibnamefont {Feist}}, \ and\ \bibinfo {author} {\bibfnamefont {F.~J.}\ \bibnamefont {Garcia-Vidal}},\ }\href {\doibase 10.1021/acs.jpcc.5b11654} {\bibfield  {journal} {\bibinfo  {journal} {J. Phys. Chem. C}\ }\textbf {\bibinfo {volume} {119}},\ \bibinfo {pages} {29132} (\bibinfo {year} {2015})}\BibitemShut {NoStop}%
\bibitem [{\citenamefont {Takele}\ \emph {et~al.}(2020)\citenamefont {Takele}, \citenamefont {Wackenhut}, \citenamefont {Piatkowski}, \citenamefont {Meixner},\ and\ \citenamefont {Waluk}}]{Takele2020}%
  \BibitemOpen
  \bibfield  {author} {\bibinfo {author} {\bibfnamefont {W.~M.}\ \bibnamefont {Takele}}, \bibinfo {author} {\bibfnamefont {F.}~\bibnamefont {Wackenhut}}, \bibinfo {author} {\bibfnamefont {L.}~\bibnamefont {Piatkowski}}, \bibinfo {author} {\bibfnamefont {A.~J.}\ \bibnamefont {Meixner}}, \ and\ \bibinfo {author} {\bibfnamefont {J.}~\bibnamefont {Waluk}},\ }\href {\doibase 10.1021/acs.jpcb.0c03815} {\bibfield  {journal} {\bibinfo  {journal} {J. Phys. Chem. B}\ }\textbf {\bibinfo {volume} {124}},\ \bibinfo {pages} {5709} (\bibinfo {year} {2020})}\BibitemShut {NoStop}%
\bibitem [{\citenamefont {Ahn}\ and\ \citenamefont {Simpkins}(2021)}]{Ahn2021}%
  \BibitemOpen
  \bibfield  {author} {\bibinfo {author} {\bibfnamefont {W.}~\bibnamefont {Ahn}}\ and\ \bibinfo {author} {\bibfnamefont {B.~S.}\ \bibnamefont {Simpkins}},\ }\href {\doibase 10.1021/acs.jpcc.0c10360} {\bibfield  {journal} {\bibinfo  {journal} {J. Phys. Chem. C}\ }\textbf {\bibinfo {volume} {125}},\ \bibinfo {pages} {830} (\bibinfo {year} {2021})}\BibitemShut {NoStop}%
\bibitem [{\citenamefont {Grechko}\ \emph {et~al.}(2018)\citenamefont {Grechko}, \citenamefont {Hasegawa}, \citenamefont {D'Angelo}, \citenamefont {Ito}, \citenamefont {Turchinovich}, \citenamefont {Nagata},\ and\ \citenamefont {Bonn}}]{grechkoCouplingIntraIntermolecular2018}%
  \BibitemOpen
  \bibfield  {author} {\bibinfo {author} {\bibfnamefont {M.}~\bibnamefont {Grechko}}, \bibinfo {author} {\bibfnamefont {T.}~\bibnamefont {Hasegawa}}, \bibinfo {author} {\bibfnamefont {F.}~\bibnamefont {D'Angelo}}, \bibinfo {author} {\bibfnamefont {H.}~\bibnamefont {Ito}}, \bibinfo {author} {\bibfnamefont {D.}~\bibnamefont {Turchinovich}}, \bibinfo {author} {\bibfnamefont {Y.}~\bibnamefont {Nagata}}, \ and\ \bibinfo {author} {\bibfnamefont {M.}~\bibnamefont {Bonn}},\ }\href {\doibase 10.1038/s41467-018-03303-y} {\bibfield  {journal} {\bibinfo  {journal} {Nat. Commun.}\ }\textbf {\bibinfo {volume} {9}},\ \bibinfo {pages} {885} (\bibinfo {year} {2018})}\BibitemShut {NoStop}%
\bibitem [{\citenamefont {Vietze}\ \emph {et~al.}(2021)\citenamefont {Vietze}, \citenamefont {Backus}, \citenamefont {Bonn},\ and\ \citenamefont {Grechko}}]{vietzeDistinguishingDifferentExcitation2021}%
  \BibitemOpen
  \bibfield  {author} {\bibinfo {author} {\bibfnamefont {L.}~\bibnamefont {Vietze}}, \bibinfo {author} {\bibfnamefont {E.~H.~G.}\ \bibnamefont {Backus}}, \bibinfo {author} {\bibfnamefont {M.}~\bibnamefont {Bonn}}, \ and\ \bibinfo {author} {\bibfnamefont {M.}~\bibnamefont {Grechko}},\ }\href {\doibase 10.1063/5.0047918} {\bibfield  {journal} {\bibinfo  {journal} {J. Chem. Phys.}\ }\textbf {\bibinfo {volume} {154}},\ \bibinfo {pages} {174201} (\bibinfo {year} {2021})}\BibitemShut {NoStop}%
\bibitem [{\citenamefont {Seliya}, \citenamefont {Bonn},\ and\ \citenamefont {Grechko}(2023)}]{SeliyaExtractingTheSampleResponse2023}%
  \BibitemOpen
  \bibfield  {author} {\bibinfo {author} {\bibfnamefont {P.}~\bibnamefont {Seliya}}, \bibinfo {author} {\bibfnamefont {M.}~\bibnamefont {Bonn}}, \ and\ \bibinfo {author} {\bibfnamefont {M.}~\bibnamefont {Grechko}},\ }\href {\doibase 10.1063/5.0138442} {\bibfield  {journal} {\bibinfo  {journal} {J. Chem. Phys.}\ }\textbf {\bibinfo {volume} {158}},\ \bibinfo {pages} {134201} (\bibinfo {year} {2023})}\BibitemShut {NoStop}%
\bibitem [{\citenamefont {Schwennicke}\ \emph {et~al.}(2025)\citenamefont {Schwennicke}, \citenamefont {Koner}, \citenamefont {P{\'{e}}rez-S{\'{a}}nchez}, \citenamefont {Xiong}, \citenamefont {Giebink}, \citenamefont {Weichman},\ and\ \citenamefont {Yuen-Zhou}}]{Schwennicke2024}%
  \BibitemOpen
  \bibfield  {author} {\bibinfo {author} {\bibfnamefont {K.}~\bibnamefont {Schwennicke}}, \bibinfo {author} {\bibfnamefont {A.}~\bibnamefont {Koner}}, \bibinfo {author} {\bibfnamefont {J.~B.}\ \bibnamefont {P{\'{e}}rez-S{\'{a}}nchez}}, \bibinfo {author} {\bibfnamefont {W.}~\bibnamefont {Xiong}}, \bibinfo {author} {\bibfnamefont {N.~C.}\ \bibnamefont {Giebink}}, \bibinfo {author} {\bibfnamefont {M.~L.}\ \bibnamefont {Weichman}}, \ and\ \bibinfo {author} {\bibfnamefont {J.}~\bibnamefont {Yuen-Zhou}},\ }\href {\doibase 10.1039/D4CS01024H} {\bibfield  {journal} {\bibinfo  {journal} {Chem. Soc. Rev.}\ }\textbf {\bibinfo {volume} {54}},\ \bibinfo {pages} {6482} (\bibinfo {year} {2025})}\BibitemShut {NoStop}%
\end{thebibliography}

%

\end{document}